\documentclass[12pt,preprint]{aastex}
\begin{document}
\title{
The mass function and distribution of velocity dispersions
for UZC  groups of galaxies
}
\author{Armando Pisani}
\affil{Astronomy Department, University of Trieste, via G.B. Tiepolo 11, I-34131 Trieste, Italy
\newline
Istituto di Istruzione Statale Classica Dante Alighieri, Scientifica Duca degli Abruzzi e Magistrale S.
Slataper, via XX settembre 11, I-34170 Gorizia, Italy}
\email{pisani@ts.astro.it}
\author{Massimo Ramella}
\affil{INAF, Osservatorio Astronomico di Trieste
\newline via G. B. Tiepolo 11, I-34131 Trieste, Italy}
\email{ramella@ts.astro.it} 
\and 
\author{Margaret J. Geller}
\affil{Smithsonian Astrophysical Observatory
\newline 60, Garden Str., Cambridge, MA02138, U.S.A.}
\email{mjg@cfa.harvard.edu}
\newcommand{\mb}{@@m-$>$}
\newcommand{\me}{@@$<$-m}
\newcommand{\be}{\begin{equation}}
\newcommand{\ee}{\end{equation}}

\newcommand{\st}{$\sigma_T \;$}
\newcommand{\sv}{$\sigma_v \;$}
\newcommand{\nst}{$n(\geq \sigma_T) \;$}
\newcommand{\nxt}{$n(\geq T_x) \;$}
\newcommand{\kms}{ \; km \, s^{-1} \;}
\newcommand{\nmem}{$N_{mem} \;$}
\newcommand{\rv}{$\rho(V) \;$}
\newcommand{\mvir}{$M_{vir} \;$}
\newcommand{\vmax}{$V_{max} \;$}
\newcommand{\pn}{$p(N) \;$}
\newcommand{\mn}{$\mu(\geq N) \;$}

\begin{abstract}

We measure the distribution of velocity dispersions of groups of galaxies
identified in the UZC catalog; we use the distribution to derive the group mass
function. We introduce a new method which makes efficient use of the entire
magnitude limited catalog.  Our determination of \nst includes a significant
contribution from low luminosity systems that would be missing in a volume
limited sample.  We start from a model for the probability density function of
the total number of group members and reproduce the observed distribution.  We
take several effects like  the local fluctuations in volume density, limited
sampling and group selection into account. We estimate the relation between
total number of members, total luminosity and true velocity dispersion.  We can
then reproduce not only the observed  distribution of \sv but also the
distributions of the number of group members, the total groups luminosity, and
virial mass.  The best fit to the data in the true velocity dispersion range
$100\; {\rm \kms} \leq \sigma_T \leq 750\; {\rm \kms}$ is a power law model with a slope
\nst $\propto \sigma_T^{-3.4^{+1.3}_{-1.6}}$  and a normalization $n(\sigma_T
\geq 750 \;{\rm  \kms}) = (1.27 \pm 0.21) \times 10^{-5} h^3 \; {\rm Mpc}^{-3}$ where
$\sigma_T$ is the true velocity dispersion of a group.  The predictions of
CDM models are consistent with the mass function we derive from this
distribution of velocity dispersions.
\end{abstract}


\section{ Introduction}

Groups of galaxies contain approximately half of the galaxies within a
magnitude limited redshift survey \citep[see e. g.][]{rgp02, rpg97, tb98, 
giu00, tuc00, am02, mer02}.  In hierarchical
structure formation theories, these abundant systems are the natural
progenitors of the galaxy clusters. Thus measurement of the distribution of the
physical parameters of groups  provides part of the fundamental basis for
understanding the large-scale structure of the universe \citep{ zg94, nkp94, 
diaf99, car00, gg00}.

The velocity dispersion of a system of galaxies is an indicator of the depth of
the potential associated with the system.  Hence the distribution of velocity
dispersions is the basis for the determination of the group mass function.
Although the relation linking the velocity dispersion with the gravitational
mass depends on the relative distribution of visible and dark mass and on the
dynamical state of the galaxy system (\citealt{diaf93}, but see also
\citealt{xfw00}), the velocity dispersion itself is a physically interesting
parameter.  The data readily provide an estimate of the group velocity
dispersion which is robust against the inclusion of faint members \citep{rgh95,
rfg96, mah99}.  The small median number of group members introduces a large
scatter in the  observational estimate, \sv, of the dispersion of a system
relative to the true value, \st.  This finite sampling effect also introduces
distortion in the distribution of \sv.

Several investigators have determined \nst for optical galaxy clusters
\citep[e. g. ][]{bah02, rb02, bor99, gir98, fad96,  maz96, bc93, biv93}.  In
addition, there are also measurements of the cluster x-ray temperature function
\citep{ike02, pier01, bla00, h00, mar98, ha91, ed90}.  On the theoretical side,
large numerical simulations provide estimates of the predicted group/cluster
mass function \citep[e. g. ][]{jen01, gov99, diaf99, lc94, thom98}.  for a
variety of cosmological models.

For galaxy groups with low luminosity and hence low velocity dispersion, some
determinations of \nst rest on small samples of systems in volume limited
catalogs \citep[e. g. ][]{zab93}. Others contain residual biases resulting from
the magnitude limit of the samples \citep[e. g. ][]{moo93, gg00, hein03}.  In
principle, restriction to a volume limited catalog mitigates the problem of
selection effects introduced by the apparent magnitude limit of the redshift
survey \citep{zab93}.  However, low luminosity systems are virtually absent
from volume limited catalogs.

The determination of \nst from magnitude limited samples requires correction
for the selection function. The standard approach is to weight each group by
the maximum volume where a group could still be recognized as a triplet. The
weighting scheme assumes that throughout the mass range the catalog covers a
fair sample of the universe.  These determinations also assume that the
mass-to-light ratio, M/L, is independent of velocity dispersion.

A reliable determination of the mass function of groups is particularly
interesting because there are still significant discrepancies between theory
and observations at the low-mass end of the distribution 
\citep{mar02, hein03}.
Standard observational approaches to the determination of \nst are most
uncertain at the low-mass end.  For example \citet{zab93} lack
low-mass systems because of the absolute magnitude cut of their volume limited
sample.  For larger samples, the assumption of constant number of groups per
unit volume is a potential problem. Even in these larger surveys, low mass
systems probe a small volume where the distribution of systems is not
homogeneous.

Here we introduce a method for constructing \nst  from all of the groups
identified in the magnitude limited UZC redshift survey. We improve the
statistics of the \nst determination  in the range $100 {\rm \kms}  \leq$ \st
$\leq 750 {\rm \kms}$ . We determine \nst in a homogeneous way throughout the
entire range in \st.  We account for the effects of limited sampling on the
estimated velocity dispersion of each group and for local fluctuations in the
number density of groups.

We demonstrate that in the UZC, the  intrinsic relations among the luminosity,
velocity dispersion and richness of groups are well-represented by power laws.
By using these model relations we can satisfactorily account for the observed
distributions of all the group parameters. These scaling relations for groups
may be useful for evaluating the evolution of group properties in catalogs
extracted from deeper surveys.

Section~3 discusses the group catalog. Section~4  outlines the main problems in
estimating \nst and the basic method we apply. Section 5 discusses a power-law
model for the relation between group richness and velocity dispersion.  Section
6 describes the entire analysis method in detail and shows how we use the
relation in Section 5.  Section 7 contains the distribution of velocity
dispersion and the group mass function. In Section~8 we discuss the scaling
relations and demonstrate the broad consistency of the method we use.  We
summarize in Section~9.

\section{ The group catalog}

We base our analysis on the UZC catalog \citep{fal99} in the region
defined by $-2.5^{\circ} \leq  \delta_{1950} \leq 50^{\circ}$ and $8^{h} \leq
\alpha_{1950} \leq 17^{h}$ in the North Galactic Cap and  by $20^{h} \leq
\alpha_{1950} \leq 4^{h}$ in the South Galactic Cap.  In this region the
completeness in redshift is $98 \%$ . We  discard the region $-13^{ \circ} \leq
b \leq 13^{\circ}$  because of the greater Galactic absorption there
\citep{pad01}.

\citet{rgp02} describe the catalog of UZC groups and list the membership of all
of the groups analyzed here.  \citet{rgp02} identify the UZC groups with a
friends-of-friends algorithm (FOFA) \citep{rpg97, hg82}. This algorithm is
currently in wide use \citep{am02, mer02, gg00, tuc00, tb98}
along with some applications of the dendogram analysis
applied by \citet{mat79} and by \citet{tul87}.  The groups  \citet{rgp02}
identify are  number density enhancements with  $\delta \rho_N / \rho_N \geq
80$ in redshift space $(\alpha,\delta,V)$.

The FOFA associates galaxies with a projected separation on the sky less than
$D_{0} = 0.25 h^{-1} \ $ Mpc and  a line-of-sight velocity difference less than
$V_{0}=350\  {\rm \kms}$ at a reference fiducial velocity $v_F = 1000\ $ km
s$^{-1}$. We scale the linking parameters $D_L = R(V) \times D_0$ and $V_L =
R(V) \times V_0$ with a function $R(V)$ according to the prescription of
\citet{hg82}.  We assume that the UZC galaxy luminosity function (LF) is in the
\citet{s76} form and adopt $M_{\star}=-19.1$, $\alpha=-1.1$, and
$\phi_{\star}=0.04\ h^3$ Mpc$^{-3}$.  We obtain these parameters by convolving
the LF determined by \citet{mar94} for a very similar sample with a Gaussian
0.3 magnitude error. These parameters improve those used by \citet{rpg97} to
identify groups within the CfA2N survey.  The differences between the
\citet{rpg97} groups and the groups we identify now within the same region are
negligible.

Within the UZC we identify 301 groups with at least 5 members and with mean
velocities $500 {\rm km s^{-1}} < V < 12000 {\rm km s^{-1}}$. At the minimum
group radial velocity $V_{min}$= 500 ${\rm \kms}$, the apparent magnitude limit
of UZC, $m_{lim} = 15.5$, corresponds to an absolute magnitude $M_{lim}=
m_{lim} -25 - 5 \log(V_{min} / H_{0}) = -13.00 + 5 \log (h)$ and, with $h = 1$,
to a luminosity $\log(L_{lim}/L_{\odot}) = 7.39$.

To eliminate 39 groups with very low intrinsic luminosity, we limit our
analysis to groups with mean velocity $V \geq 3000 $ km s$^{-1}$.  We refer to
the remaining  $N_G =262$ groups as the UZCGG sample.  These groups have true
velocity dispersions larger than $\sim 2.5$ times the errors in the individual
redshift determinations.  The lower limit in the true velocity dispersion of
our 262 groups is comparable with the lower limit of the velocity dispersion of
the x-ray emitting groups detected within UZCGG \citep{mah99}.

In a group catalog selected from a redshift survey with a FOFA, some fraction
of the groups are accidental superpositions. We have two measures of the
fraction of true physical systems in the UZCGG.  \citet{rpg97} use geometric
simulations of the large-scale structure in the northern UZC region to
demonstrate that 80\% of the groups are probably physical systems.
\citet{diaf99} compare $\Lambda$CDM simulations with the northern
portion of the UZC and conclude that the linking parameters we choose for group
selection are in the optimal range for reproducing the catalog derived from
n-body simulations. In that range the fraction of real groups is 70-80\%.
\citet{mah00} cross-correlate a large portion of the UZCGG with the ROSAT
All-Sky Survey (RASS).  61 groups in the \citet{mah99} sample have associated
extended x-ray emission. The presence of hot x-ray emitting gas bound in the
group potential well is a confirmation of the physical reality of the system.
\citet{mah00} use the groups detected in the x-ray to show that a minimum
fraction of 40\% of the groups in the UZCGG subsample are similar x-ray
emitting systems undetectable with the RASS; thus they set a lower limit of
40\% on the fraction of real physical systems in the group catalog. At least
40\% and probably 70-80\% of the UZCGG systems are real and it is thus
reasonable to use their properties to derive physical constraints on the nature
of groups of galaxies.

\section{The velocity dispersion distribution: the method}

The group velocity dispersion is a fundamental quantity both for studying the
internal dynamics of the group and for understanding the processes which
produce these galaxy systems 
\citep{fw90, moo93, wc93, zab93, cg95}. It is therefore important to have a
reliable estimate of the distribution of velocity dispersions for groups, the
most common bound systems in the universe.

Our goal is measurement of both the probability density of groups with a given
velocity dispersion, $f(\sigma_T)$, and the space density of groups as a
function of velocity dispersion, \nst.  One direct approach to the problem is
analysis of a volume limited catalog of groups. This procedure significantly
reduces the number of groups in the sample and, more importantly, it discards a
large number of low luminosity systems largely with low velocity dispersions.
For example, within the UZCGG  there are $34$ groups within the volume limited
sample with velocity limit  $V_{\star}=8300$ km s$^{-1}$, and only 20 groups
within $V_{M}=12000$ km s$^{-1}$.

Undersampling the internal velocity field of each system also affects the
determination of \nst for groups. The UZCGG groups have a median of $7$
members:  the estimate of their velocity dispersion $\sigma_v$ (corrected for
radial velocity uncertainty according to \citealt{ddt80}) has a large
scatter around the true underlying velocity dispersion, \st. We use the
unbiased estimator of \citet{led84} to compute $\sigma_v$.  To illustrate
the scatter, we use a  Monte Carlo simulation to extract $1000$ groups at
random from a specified true distribution of $\sigma_T$. We simulate the
``observed'' group velocity dispersion $\sigma_v$ under the assumption that
peculiar velocities within the group are distributed according to a Gaussian
with true dispersion $\sigma_{T}$.  The distribution of mean radial velocity
and richness of the simulated groups is the same as for the groups observed in
the UZCGG sample. Figure 1 shows the relation between the true $\sigma_T$ and
the Monte Carlo sampled $\sigma_v$ (the straight line represents equality of
the two measures).  The spread is quite large; it decreases at larger
$\sigma_v$ because the number of members observed is typically larger.  For
poorly sampled systems, the observed value of $\sigma_v$ is a poor estimator of
the true $\sigma_T$. In conclusion, the estimate of the distribution of the
velocity dispersion is severely affected by several observational and sampling
effects.

To avoid these difficulties we assume that the peculiar velocities of group
member galaxies are gaussian-distributed. This assumption is warranted by
our deeper sampling of groups with associated x-ray emission \citep{mah99}.
With an average of 35 galaxies per group, the velocity distributions are
consistent with a Gaussian. 

We  simulate the observed value of
$\sigma_v$ by  Monte Carlo-sampling of the Gaussian velocity field with
dispersion \st.  We use the data to demonstrate that there is a relation
between \st and the absolute group richness $N$.  We confirm the relation by
examining a set of 43 more deeply sampled groups from \citet{mah99}
and \citet{mg01}.  We  account for the selection effects introduced
by: a) the apparent magnitude limit of the galaxy catalog, b) the distance
dependent criterion of inclusion of a group in our analysis:  $N_{mem} \geq
N_{min}=5$, and finally c) fluctuation in the space density of groups \rv.

We select groups according to the number of members. We therefore must estimate
the probability density function, $p(N)$, of the group absolute richness $N$,
i.e. the total number of group members brighter than $L_{lim}$ (or
$M_{lim}=-13.00$).  We use a power law model for $p(N)$ and constrain both its
slope and normalization.  From $p(N)$ we predict the number density of groups
with more than $N$ members.

By using the relation  between group richness and velocity dispersion along
with the estimate of $p(N)$, we can compute both $f(\sigma_T)$ and \nst. From
these two distributions we can use Monte Carlo simulations to  predict the
observed distribution of $\sigma_v$.  We can also predict  the distribution of
observed group luminosities, $L_{mem}$, provided that the luminosity function
is universal.

The procedure includes the following steps:

\begin{description}
\item[step 1:]
we estimate the relation between the number of group members
(richness) $N$ and
the true velocity dispersion \st (Section~5)
\item[step 2:] we choose a model for the probability density function
of the group
richness $N$: $p(N)$ (Section~6)
\item[step 3] we estimate the group selection function by using
$p(N)$ (step 2) and then, by comparison with the data, we obtain
the radial distribution of group density $\rho(V)$ (Section~6)
\item[step 4:] by using the $N$ vs. $\sigma_T$ relation (step 1), the
model
for $p(N)$ (step 2) and the estimate of $\rho(V)$ (step 3),
we compute the cumulative distribution of the velocity dispersions
$n(\geq \sigma_v)$ and compare it with the real data in order to
assess the agreement with $p(N)$ from step 2 (Section~7)
\item[step 5:] by using the relations among richness, total luminosity,
\st and the group mass,
we compare the model distributions of all these quantities with the
observations (Section~8).
\end{description}

The main advantages of this procedure are; a) we determine \nst for groups with
$L_T$ an order of magnitude fainter than for a volume limited catalog of groups
identified within the same region b) we need not assume a constant space
density of groups, and c) we use a sample of groups an order of magnitude
larger than the the volume limited sample, and d) the mass function estimate
does not require the assumption of constant $M/L$.

\section {Group richness and velocity dispersion}

Several investigators have explored the relationship between various measures
of ``richness'' and the velocity dispersion of rich clusters of galaxies
\citep{bc93}: richer clusters tend to have larger velocity
dispersions.  \citet{bah88}, for example, fits a power-law relation between
between the galaxy count within 0.25 h$^{-1}$ Mpc and the system velocity
dispersion for 23 Abell clusters and a sample of groups \citep{tg76}.
More recently, \citet{yee02} have shown that the richness
of CNOC1 clusters of galaxies is well correlated with their main physical
parameters as derived from both optical and x-ray observations.
Here we use the UZCGG data along with a set of well-sampled x-ray emitting
groups \citep{mah99, mg01} to examine the relationship between group richness
and velocity dispersion.

We expect some relation between the total number of group members $N$ and the
true velocity dispersion, $\sigma_{T}$, because $N$ and $L_T$ are related
through the luminosity function, $L_T$ and mass are related for bound systems,
and mass and velocity dispersion are also related (if light traces mass).  We
do not require the same M/L for groups at every velocity dispersion.  As a
simple approximation we choose a power-law model for the relation between $n$
and $\sigma_{T}$:

\begin{equation}
N = N_{0} \left( \frac{\sigma_T}{\sigma_{0}} \right)^a
\label{N_vs_sigmat}
\end{equation}
where $N_{0}$ and $\sigma_{0}$ are scale factors and $a$ is the
exponent.

To estimate the parameters  $\sigma_{0}$ and $a$, we fix an arbitrary
normalization in number, $N_{0}=100$, and proceed in the following steps:

\begin{enumerate}
\item We select the $i-th$ group from the sample at random. This group
has
$N_{mem}(i)$ members at a mean radial velocity $V(i)$ and a total
number of galaxies brighter than
$M_{lim} = -13.0$, $N(i)=N_{mem}(i)/ \nu(V(i))$.
$\nu(V(i))$ is the fraction of galaxies
detected in a group at mean radial velocity $V(i)$:
\begin{equation} \nu(V(i)) = \frac{ \int_{L_{cut}(V(i))}^{+\infty}
\phi(x)
dx}{ \int_{L_{lim}} ^{+\infty} \phi(x) dx }
\label{n_vis}
\end{equation}
The luminosity $L_{cut}(V(i))$  corresponds to the
UZC apparent magnitude limit at the radial velocity $V(i)$ and
$\phi(L)$
is the luminosity function.

The true velocity dispersion of the group is, according to our model, \be
\sigma_T(i) = \sigma_{0} \left( \frac{N(i)}{N_{0}} \right) ^{1/a} \ee

\item We sample $N_{mem}(i)$ velocities from a Gaussian with standard
deviation
$\sigma_T(i)$ and compute the simulated velocity dispersion
$\sigma_{sim,v}(i).$

\item We repeat these steps until we obtain
the desired number of simulated groups.

\item We perform a KS test between
 the simulated distribution of $\sigma_{sim,v}$ and the distribution
of observed $\sigma_v$

\item We vary $\sigma_{0}$ and $a$ in the interval $a \in [1;3]$
and $\sigma_0 ({\rm \kms}) \in [400; 600]$ in a $21 \times 21$ grid,
and
maximize the KS significance level.
\end{enumerate}

We iterate the previous five steps ten times with 5000 groups in each iteration
and compute the median of the KS-test significance $S_{KS}$ of the comparison
between the predicted and the observed distributions.

With this procedure applied to the UZCGG, we obtain $\sigma_{0}=510 \;
{\rm \kms}$
and $a=1.8$ with median $S_{KS}(\sigma_{v})=0.44$. The $99 \%$
confidence
interval is $(1.4;2.1)$ for $a$ and $(460 \; {\rm \kms}; 580\; {\rm \kms})$ for
$\sigma_0$.
  The
value of
$\sigma_0$ is sensitive to the definition and scale of ``richness.''

Figure 2a  shows the contours of constant $S_{KS}(\sigma_{v})$ in the parameter
space. The thick outermost isopleth is the 99\% confidence level, the middle
contour traces the $90\%$ confidence level, and the inner thin contour traces
the $50 \%$ confidence level. Note that the values of $\sigma_{0}$ are scaled
for group richness estimated at $V\geq 3000 {\rm \kms}$ and hence for members
brighter than $-16.89$. The corresponding value of the best fit $\sigma_{0}$
scaled to $V\geq 500 {\rm \kms}$ and $M_{lim}=-13.00$ is $230 \;{\rm \kms}$.

To test  the model in equation eq.~\ref{N_vs_sigmat} further, we obtain
$\sigma_{0}$ and $a$ for a set of X-ray emitting groups kindly provided to us
by A. \citet{m99}. \citet{mah00} identified a set of groups with extended x-ray
emission by cross-correlating a portion of the UZC with the ROSAT All-Sky
Survey. The extended x-ray emission essentially guarantees the reality of the
system.  \citet{m99} and \citet{m03} obtained deeper redshift surveys of the
groups and increased the typical membership to 30 galaxies. We use these
enhanced data here to test the scaling relations derived for the UZC groups.

There are 43 groups in the Mahdavi sample with mean radial velocities $V\geq
5000 {\rm \kms} $ and galaxies brighter than $m\leq 16.5$. The average number
of group members is $20$ and hence the value of the velocity dispersion is
estimated more reliably than for UZCGG. Moreover, the diffuse X-ray emission
strongly reduces the false group identification problem.  Figure 2b shows the
KS-test confidence level contours for the X-ray sample. The best fit value for
the slope $a$ is $1.7$. The $99\%$ confidence intervals are $(1.0,2.7)$ for $a$
and $(410 {\rm \kms} , 590 {\rm \kms})$ for $\sigma_{0}$.  Because of the
different completeness limits and different cuts in mean radial velocity, the
values of the $\sigma_{0}$ normalization in the two plots for X-ray and UZCGG
differ by a factor 1.06. This offset improves the overlap of the contours in
the $\sigma_{0}$ direction.  Although the area covered by the contours is quite
large, it is quite remarkable that X-ray groups yield  values for $a$ and
$\sigma_0$ nearly coincident with the UZCGG sample.

Figure 3 shows N as a function of $\sigma_v$ for the x-ray sample.  The solid
line is the result we obtained with the Monte Carlo procedure.  The thin line
is the bisector of the two fits obtained from a straightforward $\chi^2$ fit to
N vs $\sigma_v$ and to $\sigma_v$ vs N respectively.  We find $\log(N)
=1.27^{+0.83}_{-0.42}\times \log(\sigma_{v}) -1.47^{+1.14} _{-2.04}$ consistent
with the Monte Carlo approach.  The Monte Carlo yields a steeper slope because
it weights less the rare groups with very low velocity dispersion which also
depart from the $L_x-\sigma$ scaling relations \citep{mg01}.

Finally we note that our result agrees with the power law relation that
\citet{yee02} find between  richness parameter, $B_{cg}$, and  radial velocity
dispersion, $\sigma_1$, for CNOC1 clusters. \citet{yee02} find $B_{cg} \propto
\sigma_{1}^{1.8\pm0.2}$ within the range $600 {\rm \kms} \leq \sigma_1 \leq
1300 {\rm \kms}$. The agreement is interesting also because there is only a
marginal overlap between our high $\sigma$ range and the low $\sigma$ range of
CNOC1 clusters.

\section{The probability density function of groups}

Here we investigate the probability density function of groups, $p(N)$, for the
UZCGG sample.  Most (92\%) of the groups have $N \leq N_r = 840$ (or \st $\leq
\sigma_r = 750 {\rm \kms}$).  Sparse sampling of the richest systems (clusters)
limits the range over which we can reliably estimate $p(N)$ to $20 < N < 840$.
In our determination of $p(N)$, we include all groups.  Groups with with $N
\geq N_r$ are visible (i.e. $N_{obs}\geq 5$) throughout the whole sample
volume. These groups have marginal weight in the determination of the slope of
$p(N)$ but they are important for  the normalization of \nst.

We assume a power-law model for  $p(N)$ :
\be
p(N) = A \left( \frac{ N}{N_{norm}} \right) ^{-\gamma}
\label{pdf_n_model}
\ee
where $N_{norm}$ is a scale factor and $A$ is a normalization constant.

Here we set $N_{norm}=N_{min}$, with $N_{min}$ = 5 (see the previous
section).
We note that the choice of $N_{norm}$ is arbitrary because the
power-law is scale free.

The \citet{ps74} formalism, PS hereafter, provides a basis for this
assumption.  \citet{vs97} argue that in
the low mass range occupied by poor galaxy systems the probability density
function $p(N)$ is  well described by a power law. Recently, \citet{jen01} 
 use numerical simulations to obtain similar results.  We make the
argument appropriate for our data here.

The PS model describes the mass
function  as:
\be
n_{PS}(M) dM = K \left( \frac{M}{M_{\star}} \right)^{\alpha}
\exp\left(- \left(\frac{M}{M_{\star}} \right)
^{\beta} \right) d \left( \frac{M}{M_{\star}} \right).
\ee
where $K$ is a normalization factor, $M_{\star}$ is a scale factor and the
exponents $\alpha$ and $\beta$ are linked to the spectral index, $n_{eff}$, of
primordial mass fluctuations: $\alpha = n_{eff}/6-3/2$ and $\beta=1+n_{eff}/3$.
If we suppose that the group mass $M$ is a power law function of the group
number of galaxies $N$, \be M \propto N^k \label{mass_number} \ee we can
rewrite the mass function in terms of $N$:

\be
 n_{PS} (N) dN = K^{\prime}
\left( \frac{N}{N_{\star}} \right)^{\alpha{k}+k-1} \exp \left( - \left(
\frac{N}{N_{\star}} \right)^{\beta{k}}\right) d \left( \frac{N}{N_{\star}}
\right) \label{psmodel} 
\ee
 The local slope of the PS model is:
 \be \gamma =
\frac{d}{d \ln(x) } \ln n_{PS}(x) \ee 
where $x=N/N_{\star}$. In other words:
\be 
\gamma = \beta{k} x^{\beta{k}} - \alpha{k}-k+1 \label{gamma_beta}
 \ee

The slope $\gamma$ depends on $x$ and hence on $N$, but for the  range of
interest here ($20 \leq N \leq 840$), $\gamma$ goes from $2.2$ to $3.3$.  This
range is in good agreement with the $90\%$ confidence interval in our fit,
namely $(2.7,3.2)$  (see section 7).  By taking  $N$ as the average value $<N>$
of the $N$ estimated for the observed groups, we can conclude that Equation
\ref{gamma_beta} links the two parameters $n_{eff}$ and $N_{\star}$ that
characterize the PS model. Given a value of $n_{eff}$ we should obtain a ``best
fit'' $N_{\star}$:

\be
N_{\star} = <N> \left( \frac{\beta{k}}{\gamma +\alpha{k}+k-1}
\right)^{1/(\beta{k})}
\label{nstar}
\ee
is shown in Figure 4.

Over the range covered by our data, the Press-Schechter model is
well-approximated by a simple power law model for the mass and $N$ functions.

We next evaluate $\gamma$ in Eq.~\ref{pdf_n_model} from the observed group
catalog.  We start by estimating the probability of finding a group at mean
velocity $V$ with at least five observed members, $N_{mem} \geq N_{min}=5$.  We
call this function the  group selection function, $\Sigma(V)$.

The probability that a group has at least $N$ total members is:
\be
S(N) = \int_{N}^{+\infty} p(x) dx
\label{S_N}
\ee
with $S(N)$ normalized so that $S(N_{norm} = N_{min} = 5) =1$.

Consequently, the group selection function
 $\Sigma(V)$ is:

\be
\Sigma(V) = S \left( \frac{N_{min}}{\nu(V)} \right)
\label{sel_V}
\ee

Clearly, $\Sigma(V)$ depends on both $\phi(L)$ and $p(N)$.

 From the group selection function, we can compute the distributions of $V$ and
$N_{mem}$ from the model for $p(N)$.  The distribution of group radial
velocities is:

\be
C(V) = \int_{V_{m}}^{V} \Delta \Omega v^{2} \rho(v) \Sigma(v) dv
\label{C_V}
\ee
where $V_m = 3000 \; {\rm \kms}$,  $\Delta \Omega = 3.16$ sr is the
solid
angle of UZCGG, and $\rho(V)$ is the density of groups as a function of
their
mean radial velocity.  The distribution of observed members is:

\be
Q(N_{mem}) = \int_{V_{m}}^{V_M}
 \Delta \Omega v^{2} \rho(v) S \left( \frac{N_{mem}}{\nu(v)}\right) dv
\label{Q_n}
\ee
with $V_M =12000 \; {\rm \kms}$, the redshift limit of the group
catalog.

We compare the simulated  $C(V)$ and $Q(N_{mem})$ with the observed
distributions of the same quantities.  To make the comparison we must solve a
system of two equations in the two functions $p(N)$ and $\rho(V)$. We have an
assumed form for $p(N)$.  To estimate $\rho(V)$ we invert the equation for
$C(V)$ (Eq.~\ref{C_V}).  We approximate $\rho(V)$ with a sequence of
$N_{shell}$ radial velocity shells, each with constant density, i.e.:

\be
\rho(V) = \rho_{i} \; if \; V_{i-1} < V \leq V_{i}
\label{beta-shells}
\ee
with $i=1,\ldots,N_{shell}$. If there are
$\delta N_{i}$ observed groups
within the $i-th$ shell,  $\rho_{i}$ is:

\be
\rho_{i} = \frac{ \delta N_{i}}{ \Delta \Omega \int_{V_{i-1}} ^{V_{i}}
 v^{2} \Sigma(v) dv }
\ee

We choose the limits of the shells,  $V_{i}$, so that all shells contain the
same number of groups.  We start with $N_{shell}=1$  and perform a KS-test to
estimate the significance $S_{KS}(V)$ of the agreement between the model
prediction  $C(V)$ and the observed distribution of $V$. We increase
$N_{shell}$ until $S_{KS}(V) \geq 0.5$.

We have thus derived a reasonable approximation to $\rho(V)$ and we next use it
in eq.~\ref{Q_n} to compute $Q(N_{mem})$. We use a KS-test to evaluate the
agreement between $Q(N_{mem})$ and the observed distribution of \nmem.  If the
agreement is satisfactory, we conclude that we have a satisfactory slope for
$p(N)$. If, on the contrary, the distributions are inconsistent, we repeat the
whole procedure with a different slope for $p(N)$.  Once we determine \rv and
\pn, we can compute the total abundance of seen and unseen groups richer than a
fixed threshold within the UZCGG volume, i.e. the group multiplicity function
$\mu(\geq N)$.

The total number ${\cal N}_G$ of groups within UZCGG is:

\be
{\cal N}_G = \Delta \Omega \int _{V_{min}}^{V_{max}} v^{2}
 \rho(v) dv
\label{cal_G}
\ee
and the group average density is:

\be
\bar{\rho} = \frac{ {\cal N}_G}{ {\cal V}}
\ee
where
%
%
${\cal V}=1.79\times 10^{6} \; h^{-3} \; {\rm Mpc}^{3}$
is the UZCGG volume. Under the assumption that $p(N)$ does not depend
on  position within the UZCGG volume, the multiplicity function is:

\be
\mu(\geq N) = \bar{\rho} \int_{N}^{+\infty} p(x) dx
\label{mu_N}
\ee

For groups with $N$ large enough to be observable throughout the UZCGG volume,
we can compare our model prediction $\mu(N)$ directly with the observations.
We call these groups {\bf robust} and use them in order to normalize the
function $\mu(\geq N)$.

To go from \pn and \mn to \nst we use the relation between $N$ and \st
(equation \ref{N_vs_sigmat}). The data constrain the power-law slope $a$ and
the normalization $\sigma_0$.  The relations we need are:

\be
f(\sigma_T) = p(N) \frac{dN}{d\sigma_T}
\label{pdf_st}
\ee
and
\be
n(\geq \sigma_T) = \bar{\rho} \int_{\sigma_T}^{+\infty} f(x) dx
\label{n_st}
\ee

We can compute \nst by inserting equation \ref{pdf_n_model} into
equation
\ref{pdf_st}.

Because $\sigma_T$ is related to $N$ as in the eq.~\ref{N_vs_sigmat},
we can write
\be
f(\sigma_T) = \left( \frac{ \sigma_T}{\sigma_{norm}} \right)
^{-\gamma_s}
\label{pdf_st_model}
\ee
with
$\gamma_s=\gamma a - a +1$ and  $\sigma_{norm}$  is a normalization
scale.

To estimate the value of the exponent $\gamma$ we compare the observations with
the predictions of the \pn model obtained  by inserting eq.~\ref{pdf_n_model}
for $p(N)$ into eq.s~\ref{C_V} and \ref{Q_n}.  The best fit value is
$\gamma=2.9$ with a $90 \%$ confidence level interval $(2.7;3.2)$.  Figure 5
shows the KS significance level $S_{KS}$ vs $\gamma$. The curve is very well
behaved.  This fit requires $N_{shell}=7$ homogeneous density shells (see
eq.~\ref{beta-shells}). We show the density function $\rho(V)$ for the 7 shells
in Figure 6; the density rises steeply at large V (last two shells). If we
remove these shells from the analysis, the results do not change significantly.
With  $\gamma = 2.9$ our model is consistent with both the observed cumulative
distribution of $N_{mem}$  and the radial velocity distribution with
significance levels of $S_{KS }(N_{mem}) = 0.73$ and $S_{ks}(V) = 0.62$
respectively.

\section{The distribution of velocity dispersions and the
group mass function}

To compare with models we derive the velocity dispersion distribution and the
mass function.  \citet{diaf99} make an exhaustive comparison of groups catalogs
derived from the $\tau$CDM and $\Lambda$CDM simulations by \citet{kau99} with
catalogs derived from CfA2N, subset of the UZC. They use the same FOFA we
employ and vary the linking parameters over a wide range.  They also test the
sensitivity of their results to variations in the luminosity function and to the
presence of structures like the Great Wall.  They demonstrate that for the
linking parameters we use here, the catalog derived from the data agree
reasonably well with the catalogs derived from $\Lambda$ CDM provided that the
luminosity density is the same in both cases. They show that the normalization
of the distribution of velocity dispersions (and of the mass function) is much
more sensitive to the luminosity density than to variation in the linking
parameters in the group finding algorithm over a reasonable range. They
emphasize that the Great Wall plays an important role in biasing the
distribution of velocity dispersions and the mass function toward higher values
relative to the simulations. We see a similar effect and comment further below.

In the previous section we obtain the best fit value $\gamma=2.9$ (90\%
c.l. (2.7;3.2)) for the exponent of the power-law probability density function
of groups, $p(N)$.  We conclude that, at high confidence level, the density of
galaxy systems with true velocity dispersion larger than $\sigma_T$ is:

\be n(\geq \sigma_T) = (1.27 \pm 0.21) \times 10^{-5} h^3 \; {\rm Mpc}^{-3}
\left(
\frac{\sigma_T}{750 \; {\rm \kms}} \right) ^{-3.4^{+1.3}_{-1.6}}
\label{n_st_result} \ee

This distribution applies over the range $100 {\rm \kms}  \leq \sigma \leq 750 {\rm \kms}$.
The volume covered by the group catalog is too small to include many rich
clusters.  Within the UZCGG there are 20  (robust) groups visible throughout the
entire UZCGG volume. These groups have $N \geq N_{r} = 840$ corresponding to a
true velocity dispersion
$\sigma_r = 750 \; {\rm \kms}$ (see eq.~\ref{N_vs_sigmat}), the actual
upper limit of the range of $N$ (and \st) over which we
determine  \pn. For the high-\st systems our analysis does
not provide an estimate of the slope of \pn, but
it does provides an estimate of the
abundance of these systems: $\mu(N_r) = 1.27 \times 10^{-5} h^3 \; {\rm Mpc}^{-3}$.
We compute the Poisson uncertainty in $\mu_{obs}(N_{r})$ 
in the number of observed groups $N_G$:  $\delta \mu_{obs}(N_{r}) =
0.25 \times 10^{-5} h^3 \; {\rm Mpc}^{-3}$ and conclude that our prediction with the
observed value $\mu_{obs}(N_r)$ 
to within $0.68 \times \delta \mu_{obs}(N_r)$.

We also compare our determination of the abundance of massive systems with previous surveys.
Direct comparison is possible with \citet{maz96, fad96, zab93}.
These authors give \nst for clusters.  Their system abundances at $\sigma = 750
 {\rm \kms}$, are 0.2, 0.6, and 0.6  $\times 10^{-5} h^{3} {\rm Mpc}^{-3}$ respectively. Our system 
abundance is $n(\geq 750) = 1.3 \times 10^{-5} h^{3} {\rm Mpc}^{-3}$. Internal errors are of the 
order of few tenths in units of $10^{-5}$. These errors reflect the size of the samples and
 may be less important than the systematic errors introduced by various selection effects.  
 To examine the discrepancy between our estimate of $n(\geq 750)$ and previous ones, we 
summarize the characteristics of the different samples.  

The ENACS clusters \citep{maz96}
 lead to the lowest abundance of systems.  ENACS is a sample of Abell clusters selected 
according to a richness criterion. Abell clusters are an incomplete set of systems 
\citep[see e. g.][DPOSS2]{gal03}.
 More importantly, selection according to Abell richness produces a
 biased sample of velocity dispersions. \citet{maz96} are aware of this problem and 
discuss it in detail. The velocity dispersions of clusters selected according to
 richness are biased high because $\sigma$ increases with
richness and because there is a broad 
scatter around the mean relation.  \citet{maz96} use reasonable arguments to identify a
 threshold, $\sigma = 800 {\rm \kms}$, above which their cluster catalog is unbiased. In the end, 
however, unbiasedness remains an assumption.  The abundance derived by \citet{fad96}  is
 a factor of three larger than \citet{maz96}.  \citet{fad96} add poorer systems 
to the ENACS sample. In order to have a $\sigma$ distribution that takes into account poor 
systems, \citet{fad96} scale their richness distribution to mimic that of the 
Edinburgh-Durham Cluster Catalog \citep[EDCC:][]{edcc}.
 EDCC is probably a more complete catalog than 
Abell-ACO, especially at the low richness end.  \citet{fad96} perform 10,000 random 
samplings of the  velocity dispersions of their systems, constraining the richness 
distribution  of the bootstrapped sample to be the same as that of EDCC systems.  
This analysis leads to the same cluster abundance as \citet{zab93}, who use 
a combined sample of Abell clusters and the thirty densest groups selected within the 
first two slices of the CfA2N redshift survey \citep{rgh89}.

We  select systems in redshift space from a complete magnitude limited 
galaxy catalog similar to but more extensive than the one used by \citet{zab93}.
 In contrast with \citet{zab93}, we include relatively poor systems that would 
enter neither a volume limited sample nor a 2D-selected sample.  In fact, because of the 
large scatter around the mean relation between $\sigma$ and richness, several of these 
relatively poor systems have velocity dispersions $\sigma \geq 750)$.  
It is therefore not
surprising that we find a higher abundance of systems  above this threshold.

Other optical and x-ray studies have returned lower system abundances.
\citet{bc93}, for example, argue that
the abundance of clusters with velocity dispersion larger than $750 {\rm \kms}$ is
$0.2 \times 10^{-5} h^{3} {\rm Mpc}^{-3}$, much lower than our estimate.
They use a
richness limited sample of Abell clusters and obtain a scaling
relation $M\propto \sigma^{2}$ significantly different from ours
(see Table 1).

The abundance of clusters can also be estimated from x-ray selected
samples. The comparison of our abundance of systems with that of x-ray clusters
requires a relation between x-ray temperature and velocity dispersion and/or
mass. The scatter around this relation is the source of considerable
uncertainty.  \citet{rb02}  have carefully analyzed
the abundance of clusters derived from x-ray data.  
In the range $2.55\times 10^{14}\; M_{\odot} \leq M
\leq 4.4 \times 10^{14}\; M_{\odot}$ the cluster abundance is
$1.4 \times 10^{-6} \; h^{3}\; {\rm Mpc}^{-3} \leq n(\geq M) \leq 
7.0 \times 10^{-6} \; h^{3}\; {\rm Mpc}^{-3}$.
They conclude that the abundance of x-ray clusters is generally lower 
by a factor of $1.2-6.2$ than that of optically selected clusters 
\citep{gir98}. The 
origin of the discrepancy may be
either intrinsic --optical and X-ray clusters may belong to different
populations-- or may result from observational biases affecting
the mass estimates.

We may overestimate $n(\geq
750)$ somewhat because of (1) projection effects in redshift space,
(2) large-scale inhomogeneities in the galaxy distribution,
and (3) systematic errors in catalog magnitude limits.
N-body and geometrical simulations
indicate that 20\% of the 5-member groups we study
are accidental superpositions. X-ray observations \citep{mah00}
indicate that at least 40\% of the groups we study are true physical
systems.  It is difficult to correct our \nst for the presence of
false groups since we know their abundance
nor their velocity dispersion distribution.
Whatever the correction, our $n(\geq 750)$ would decrease to come into
closer agreement with the estimates of \citet{zab93}
and \citet{fad96}. 

Another possible effect driving our abundance toward a high value is the
possible over-luminosity of the UZC catalog from which we derive our group
catalog \citep{rgp02}.
The Zwicky magnitude system may be deeper than implied by its formal
magnitude limit. In this case, we detect more systems than we should
within the surveyed volume of the Universe.  The 
density of systems could be too high by as
much as a factor two. Correction for this effect would bring our estimate of
massive systems into close agreement with \citet{fad96} and with
\citet{zab93}. 

Finally, it is possible that the region of the UZC is overdense. The abundance
of groups is proportional to the galaxy density across many different surveys
\citep[see e. g.][]{ram99}.  Variations in the density on the scale of the UZC
may be as large as a factor of 2, but not more.  Further, \citet{diaf99}
emphasize that the presence of the Great Wall in the northern portion of the
UZC biases the groups sample toward higher luminosity and velocity dispersion
relative to a typical regions extracted from a $\Lambda$CDM simulation.  The
groups in the Great Wall are at a larger velocity ($\sim 8000 {\rm \kms)}$ than the
median groups redshift ($\sim 6000 {\rm \kms}$) in a simulated region. These effects
may contribute substantially to the high normalization of the mass function we
derive.

We next review estimates of the mass function of groups.  \citet{gg00} measure
the mass function of a sample of groups.  A direct comparison between their
masses and ours is not straightforward because their masses are obtained in a
model dependent way. Taken at face value (see Fig. 9), their mass function is
not significantly different from ours. Our estimate, however, extends toward
lower masses by almost an order of magnitude.

\citet{hein03} estimate the mass function of the Las Campanas Redshift Survey
groups \citep{tuc00}. The amplitude of their mass function is very low (less
than $0.1 \times 10^{-5} h^{3} {\rm Mpc}^{-3}$ for masses roughly corresponding
to $\sigma = 750 {\rm \kms}$.  \citet{hein03} compare their mass function with
the results of \citet{gg00} and with N-body simulations (see Fig.7 in
\citealt{hein03}).  The comparison demonstrates the generally low abundance  of
systems  and the marked flattening of the LCRS group mass function with respect
to the simulated mass function at the low mass end.  We obtain a similar
flattening when we do not take the large-scale variation of group number
density into account.  Low mass groups are undersampled and a V/V$_{max}$
weighting scheme does not fully recover them. 

\citet{mar02} analyze a 2dF sample of groups \citep{mer02}.
 In this case, as the
authors recognize, the flattening of the mass function at low
mass results from the low number cut-off imposed on their group
identification procedure. \citet{mar02} compare their mass function
with N-body simulations \citep{jen01} and claim  good
agreement.
 Because their best fit agrees with the 
$\Lambda$CDM model, their result is marginally consistent 
($\sim 95 \%$) with ours. 

In Figure 7 we plot \nst and its 95\% confidence level corridor.  There we also
plot the velocity dispersion distribution obtained by \citet{zab93} (long dash). 
As expected, the \citet{zab93} volume limited sample does
not include low velocity dispersions systems.

By using the relations between richness and mass (see Eq.~\ref{mass_number} and
Table 1), and between richness and velocity dispersion (eq.~\ref{N_vs_sigmat})
we can convert \nst into an estimate of the mass function.
We compare this determination with the simulations of
\citet[][Virgo Consortium]{jen01} who
derive predictions for the mass function of systems of galaxies over our
observed mass range.
Because the theoretical mass functions predicted by \citet{jen01} 
apply to masses of critical overdensities, $(\delta\rho_c/\rho)_m$, we scale
their masses to match our group number overdensity threshold
$(\delta\rho_c/\rho)_N = 80$.
We perform the scaling as in \citet{bor99} for both of the
cosmological
models in \citet{jen01}, $\Lambda$CDM ($\Omega_m = 0.3,
\Omega_\Lambda =
0.7, \sigma_8 = 0.9$) and $\tau$CDM with  ($\Omega_m = 1,
\Omega_\Lambda = 0,
\sigma_8 = 0.6$).

Figure 8 compares our power law determination of the mass function with the
theoretical results of \citet{jen01}.  We plot the cumulative mass
function $n(\geq M)$ versus the mass $M$. The solid line represents our power
law estimate, the dotted lines are 95\% confidence levels. The $\tau$CDM mass
functions lies within the 95\% confidence level over the mass range.  The mass
function predicted by the $\Lambda$CDM cosmology is only marginally consistent
with our observations. 
If we correct the mass function normalization for the presence of unphysical 
groups, the agreement with the $\Lambda$CDM model improves.

The agreement of our observations with the theoretical mass function is
remarkable. The use of our  weighting procedure is critical in obtaining this
result. In fact, if we use the standard approach and weight each group by its
maximum accessible volume, V/V$_{max}$, we obtain the thin long-dashed curve in
Figure 9. This curve shows the characteristic bending to a shallower slope at
low masses.  As discussed above, this shallower slope appears in other observed mass
functions 
\citep[e.g.:][]{mar02, gg00}.
Our procedure extends the determination of
the  mass function to masses almost one order of magnitude lower than previous
estimates.

\section{Consistency tests and scaling relations}

The mass function we derive depends on the estimate of the two functions
$\rho(V)$ and \pn. We now show that, starting from these two functions, we can
reproduce the observed distributions of all the main physical parameters of
groups, i.e. $V$, $N_{mem}$, $L_{mem}$, \sv, and virial mass, \mvir.

We proceed in the following way: we randomly sample $\rho(V)$ and generate a
group with $N$ members brighter than $L_{lim}$. We estimate the number of
visible members, \nmem, as $N_{mem}=N\cdot \nu(V)$, according to \ref{n_vis}.
We discard the group if \nmem is less than 5. If we have at least 5 members, we
estimate $L_{mem}$ by randomly sampling the Schechter (1976) LF  \nmem times.
Finally, we estimate \sv by randomly sampling a gaussian with dispersion \st,
where we compute \st  from the best fit model equation \ref{N_vs_sigmat}. We
could estimate the virial mass by using the true velocity dispersion \st:
$M_{T}=3 \sigma_T^2 R_{vir}/G$. However, in the usual analysis of an observed
group catalog, the value of \st is not available and the standard estimate of
the virial mass is $M_{vir}=3 \sigma_v^2 R_{vir}/G$. We simulate the
observationally derived value of $M_{vir}$ in the following way. We compute the
scaling relations between, for example, $N$ and $M_T$ by using the bisector of
the two straight lines we obtain first by least squares of $\log(M_T)$ versus
$\log(N)$ and second by least squares of $\log(N)$ versus $\log(M_T)$.  We
compute the simulated value of $M_{vir}$ as $M_{vir}=M_T
(\sigma_v/\sigma_T)^2$.

We stop the previous Monte Carlo simulations when we reach a total of 1000
observable groups. We compare the simulated to the observed distributions and
compute the significance of the agreement between the two distributions with a
KS-test. We repeat the whole procedure 10 times and take the median KS
significance level as a measure of the agreement between the distributions.

We obtain the following median values of $S_{KS}$: 0.43, 0.76, 0.29, 0.15, and
0.34 for $V$, \nmem, $L_{mem}$, \sv, and \mvir respectively.  Figure 10 shows
the distributions of the physical parameters of 1000 simulated groups (thin
line) together with the corresponding distributions for the observed groups
(thick line). The agreement between the observations and the predictions of the
simulations indicates that our model is a satisfactory representation of the
data.

To test the sensitivity of the exponent $\gamma$ in the \pn model, to the
parameters of the $N$ vs. \st law, we  use the most extreme values of the
parameters $a$ and $N_{0}$ in Eq.~\ref{N_vs_sigmat} within the $99 \%$
confidence level contour (see Figure 2a).  With these parameter values it is
impossible to recover both the observed distributions of the group members
$N_{mem}$ and velocity dispersion $\sigma_{v}$.

Table 1 lists the scaling relations we obtain. 
We determine the scaling relations as the angular coefficient and intercept
of the bisector of the two straight line fits $X$ versus $Y$ and 
$Y$ versus $X$,  with  $X$ and $Y$ any two related quantities in Table 1. 
We use the results of the two fits to characterize the uncertainty 
in the scaling relations. 
  
We actually use the relation between  mass, $M$, and richness, $N$, 
(Eq.~\ref{mass_number}) or, equivalently, between  $M$ and the 
velocity dispersion, $\sigma$,  to translate \nst into a mass
function, $n(\geq M)$. We also use this relation 
to recover the observed distribution of the virial mass $M_{vir}$.

Once we establish the relation between $M$ and $N$,
we also have a relation between $M$ and luminosity, $L$. As
expected, the slope of $\log(M)$ vs. $\log(N)$ and the slope of $\log(M)$
vs. $\log(L)$ are indistinguishable.
\citet{gir02} analyze the relation between  $M$ and $L$ 
for a different sample of groups and clusters.
They obtain $M \propto L^{1.34 \pm 0.03}$,  
within about one sigma of our relation. 

N-body simulations and semianalytic modeling of galaxy
formation performed by \citet{kau99} and \citet{ben00}
 gave a reliable prediction of the
$M/L$ ratio for galaxy systems and also provide an estimate of
the dependence of mass-to-light ratio, $M/L$, on $L$ (or $M$).
It is generally found that the $M\L$ ratio increases from poor
to rich systems, and eventually flattens on large scales.

\citet{yee02} measure the relation between mass and 
richness ($B_{cg}$) within the CNOC1 sample of clusters of galaxies. These
authors find  $M \propto B_{cg}^{1.64\pm0.28}$, in very
good agreement with our result. Because the intercept of this relation depends
on the magnitude completeness limit, we can compare only slopes and not
intercepts. 

As a consistency check, we also compute the relation between
$\log(L_{T})$ and $\log(\sigma_{T})$. Reassuringly, the slope
of this scaling law is the same as our best fit of the power law 
model in Eq.~\ref{N_vs_sigmat}, $N$ vs. \st.

Finally, for the sake of completeness, in Table 1 we list the relations
between $\log(M)$ vs. $\log(\sigma_{T})$, $\log(N)$ vs. $\log(R_{vir})$
and $\log(\sigma_{T})$ vs. $\log(R_{vir})$. \citet{yee02}
 fit  the scaling between the virial radius and 
richness and find $B_{cg} \propto R^{2.1\pm0.7}$,
in very good agreement with our result.
Our scaling relations significantly extend the 
data range spanned by cluster analyses. These scaling relations 
are remarkably consistent with those derived for clusters.

We also explore different forms for $p(N)$. For example, we consider a $p(N)$
that gives an exponential law for $f(\sigma_T)$, i.e. the law used  by
\citet{zab93}. This form for $p(N)$ fails to reproduce the space
density of robust groups $\mu_{obs}(\geq N_r)$ and leads to a discordant
distribution of $L_{mem}$. Another model we test is the PS
 model.  The problem with this model (Equation~\ref{psmodel}) is that it
has two free parameters, $n_{eff}$ and $N_{\star}$: our procedure can constrain
only  one parameter (see section 6). However, we can test the relation between
$n_{eff}$ and $N_{\star}$ if \pn can be approximated by a power-law (see
Equation~\ref{nstar}). We compare the PS model for \pn with our data using a KS
test. In Figure~4 we plot the 10\%, 50\%, and 99\% confidence level contours in
the $n_{eff}$ --- $N_{\star}$ plane. In the same plane we also plot $N_{\star}$
as a function of $n_{eff}$ according to Equation~\ref{nstar} where we assume
$<N> = 300$, close to the average value for our sample. For $\gamma$ we use our
best fit value, $\gamma = 2.9$. Figure~4 clearly shows good agreement between
the power-law approximation and the PS model (note that the rejection region is
external to the outermost contour (c.l. 99\%)).

\citet{sat00}   determine $n_{eff} = -1.2\pm0.3$ for masses in the range
$10^{12} -- 10^{15} M_\odot$ based on the analysis of an ASCA x-ray cluster
sample.  \citet{zan01} find similar results for these scales.  If we
use the value  $n_{eff} = -1.2$ in Equation~\ref{psmodel}, we obtain a PS model
with $N_\star \simeq 280$ corresponding to $M_\star = 1.6 \times 10^{14} \;
h^{-1}\; M_\odot$ ( $95\%$ confidence interval goes from $6.5\times 10^{13} \;
h^{-1}\; M_\odot$ to $3.6\times 10^{14} \; h^{-1}\; M_\odot$).  This value of
$M_\star$ is in reasonable agreement with the determinations of 
\citet{bc93}, namely $M_\star = (1.8 \pm 0.3) \times 10^{14} \; h^{-1}\; M_\odot$ and
\citet{gir98}: $2.6^{+0.8}_{-0.6} \times 10^{14} \; h^{-1}\; M_\odot$.
This result indicates that our analysis of  low-mass
systems of galaxies leads to an estimate of the mass function parameters 
 consistent with the results determined in the higher mass range of galaxy
clusters.

 \section{Summary}  
We measure the probability density function for the velocity dispersion of groups
 in the UZC galaxy catalog. Our
 method a) does not require that the
 group spatial density is constant, b) predicts the observed distributions of all the
 fundamental quantities of groups, i.e radial velocity, number of members, luminosity,
 velocity dispersion, and virial mass, and c) includes low luminosity systems that would
 be neglected with the usual analysis of a volume limited sample.  The best fit for the 
slope of \nst is $-3.4 \;(-2.6;-4.7)$ over the range $100 {\rm \kms}  \leq \sigma \leq 750 {\rm \kms}$.
Over this range,  the Press-Schechter model is well-approximated by a simple 
power law model.   Our model predictions agree quite well with the predictions of \citet{jen01}
 for the  CDM models. 

Our weighting procedure is critical in 
the mass function determination. If we use the standard approach and weight each group by
 its maximum accessible volume, V/V$_{max}$, we obtain the characteristic bending of the
 mass function to a shallower slope at low masses. This shallower slope appears in other
 observed mass functions \citep[e. g.][]{mar02, gg00}.  Our 
procedure extends the determination of the  mass function to masses almost one order of 
magnitude lower than previous estimates.      

For massive systems, the density of groups
with true velocity dispersion larger than $750 \; {\rm \kms}$ 
is $(1.27 \pm 0.21 ) \times 10^{-5} h^3 \; {\rm Mpc}^{-3}$.  Our determination is larger by a 
factor of two than previous determinations. We examine this discrepancy and identify several
 possible sources for this discrepancy: a) incompleteness of cluster catalogs used in previous
 works, b) presence of false groups in our catalog produced by projection effects in 
redshift-space, c) systematic errors of the Zwicky magnitude system (may be deeper than 
its nominal magnitude limit would imply), d) real overdensity of the region of the UZC. 
Corrections for any of these effects would bring the abundance determinations into closer
 agreement.  

The mass function we derive depends on the estimate of the two functions 
$\rho(V)$ and \pn. Starting from these two functions, we reproduce the observed distributions 
of all the main physical parameters of groups, i.e. $V$, $N_{mem}$, $L_{mem}$, \sv, and
 virial mass, \mvir.  We also obtain scaling relations between these physical parameters. 
Our scaling relations significantly extend the data range spanned by cluster analyses and 
 are remarkably consistent with those derived for clusters. 

\section{Acknowledgments} We benefited from extensive discussions with 
Antonaldo Diaferio about the comparison between observed and simulated 
group catalogs.
We thank Andisheh Mahdavi for providing us with the data of x-ray emitting 
groups.
We also thank Adrian Jenkins for his help with the CDM mass functions.
We thank the anonymous referee for helpful suggestions. 
This work was partially supported by the Italian Space Agency (ASI) and by 
Italian Ministry of Education, University and Research (MIUR grant COFIN 
2001028932).



\begin{figure}
\figurenum{1}
\plotone{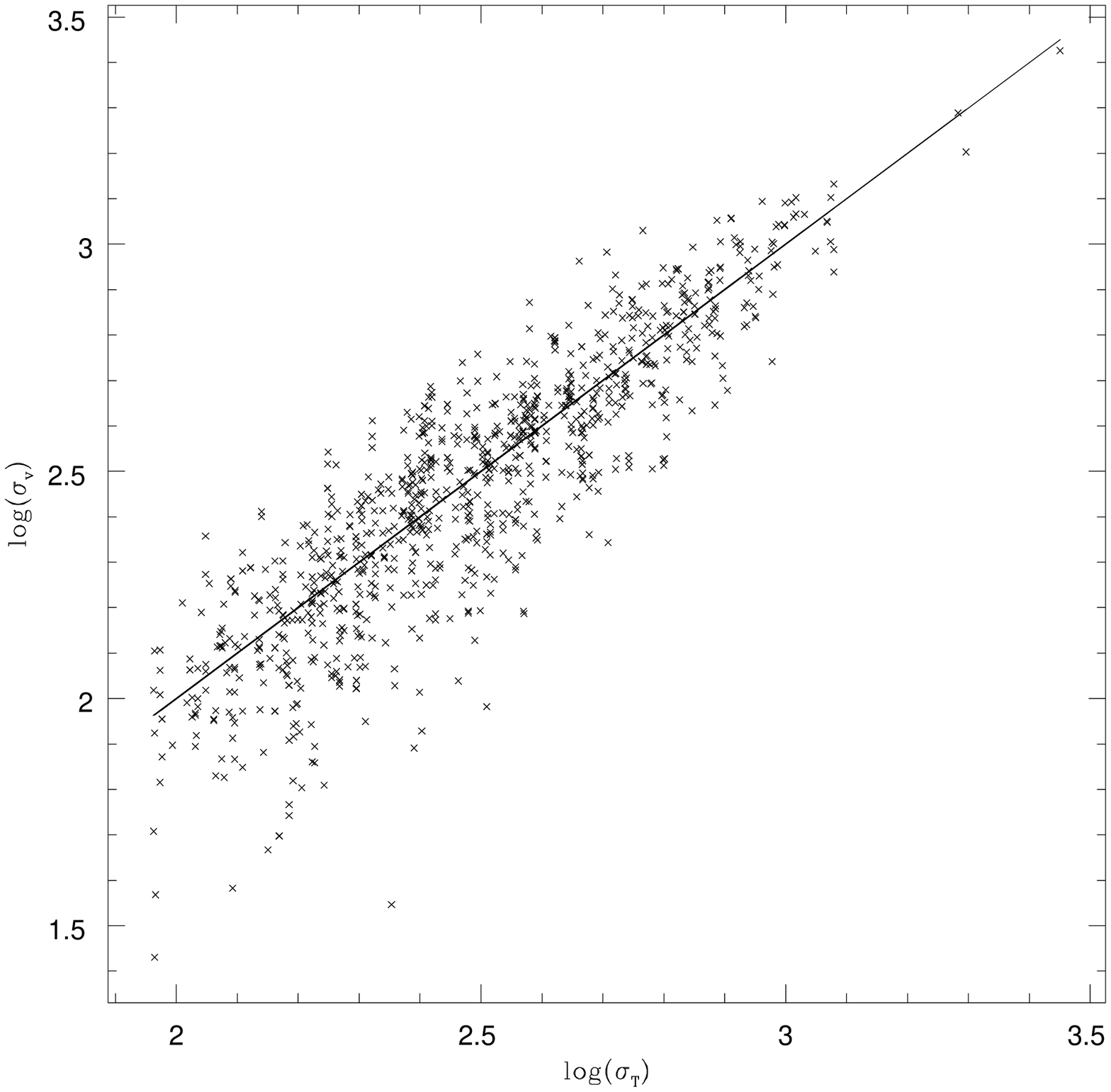}
\caption{The relation between the true velocity dispersion
 $\sigma_T$  and
observed velocity dispersion $\sigma_v$ for a Monte Carlo
simulation. The solid line represents the identity.}
\end{figure}

\begin{figure}\figurenum{2a}
\plotone{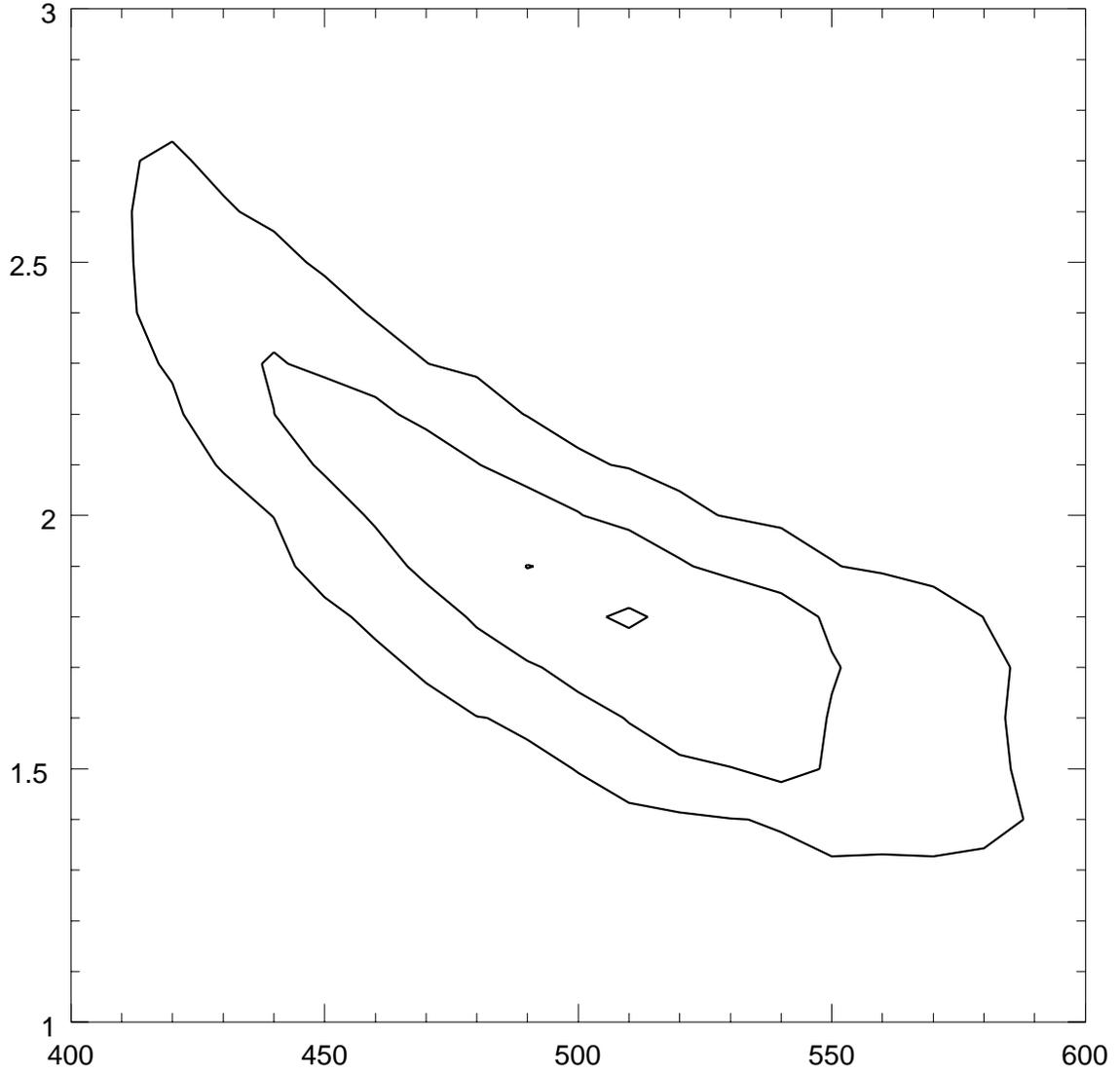}
\caption{The contour levels of the model
$N$ versus $\sigma_T$ (eq.~\ref{N_vs_sigmat})
 for the UZCGG groups with 5 or more members. The outer thick
contour
 traces the $99 \%$ confidence region, the middle contour traces the
$90\%$ confidence level, and
the inner thin contour traces
the
$50 \%$ confidence level.}
\end{figure}

\begin{figure}\figurenum{2b}
\plotone{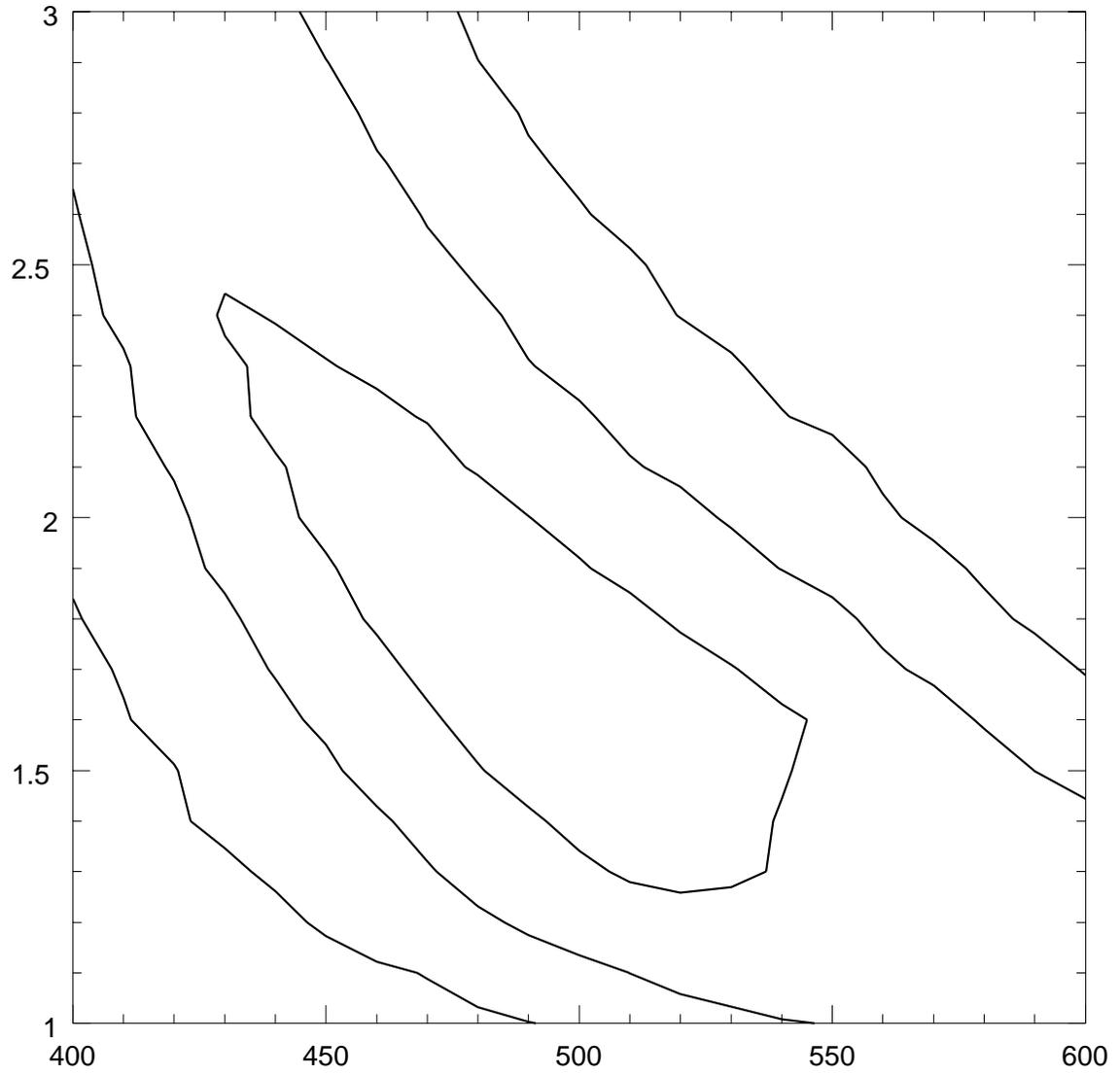}
\caption{Same as Figure 2a for x-ray emitting groups
}
\end{figure}

\begin{figure}\figurenum{3}
\plotone{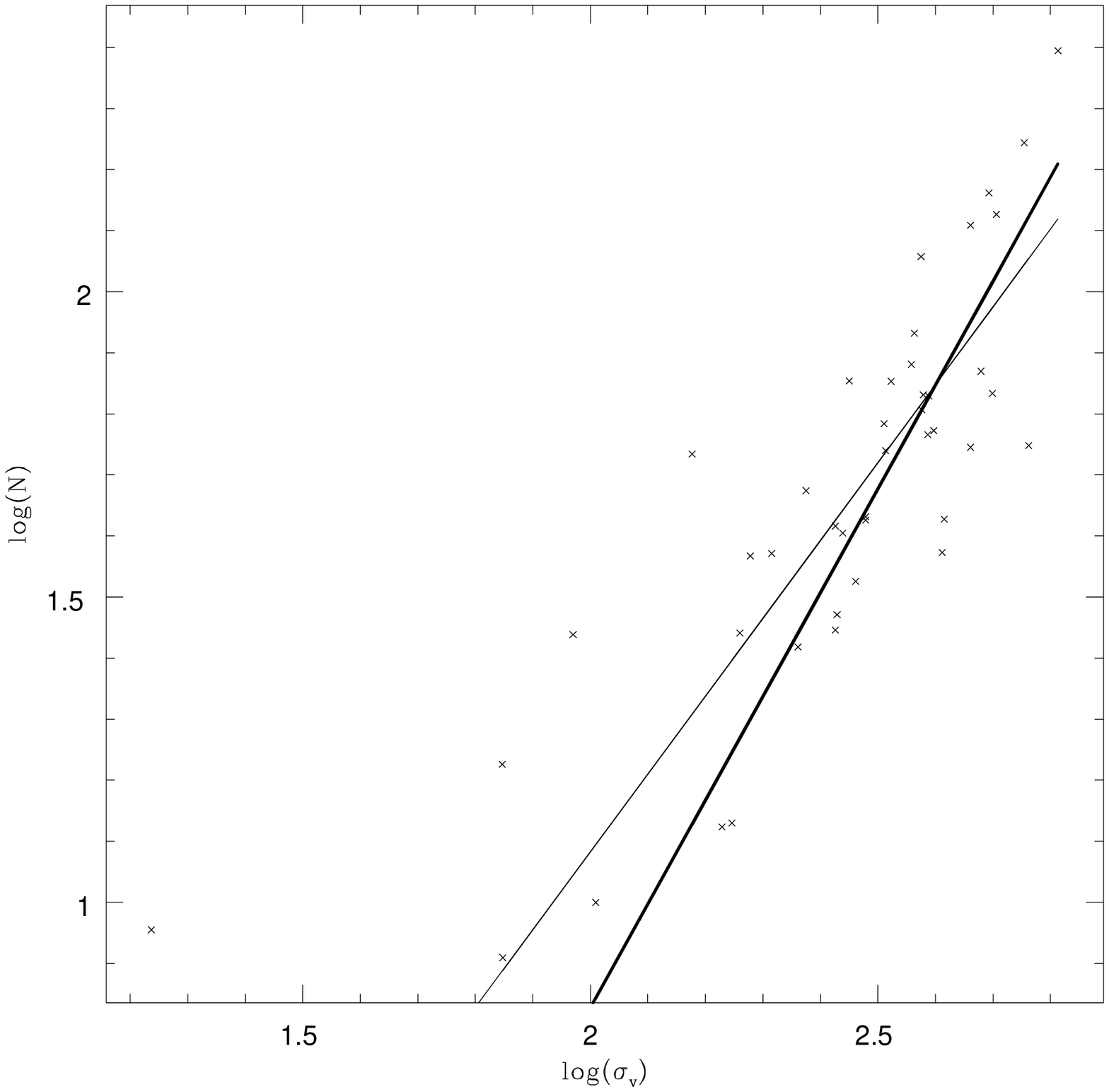}
\caption{The relation between the group richness $N$ and the
velocity
dispersion $\sigma_{v}$ for the x-ray emitting groups. The
thin line
results from the bisector of the two least squares straight lines (one by
fitting $N$ vs.
$\sigma_{v}$ and the other by fitting $\sigma_{v}$ vs. $N$), the thick
line shows that
our Monte Carlo approach has less sensitivity to low
$\sigma_{v}$ groups.}
\end{figure}

\begin{figure}\figurenum{4}
\plotone{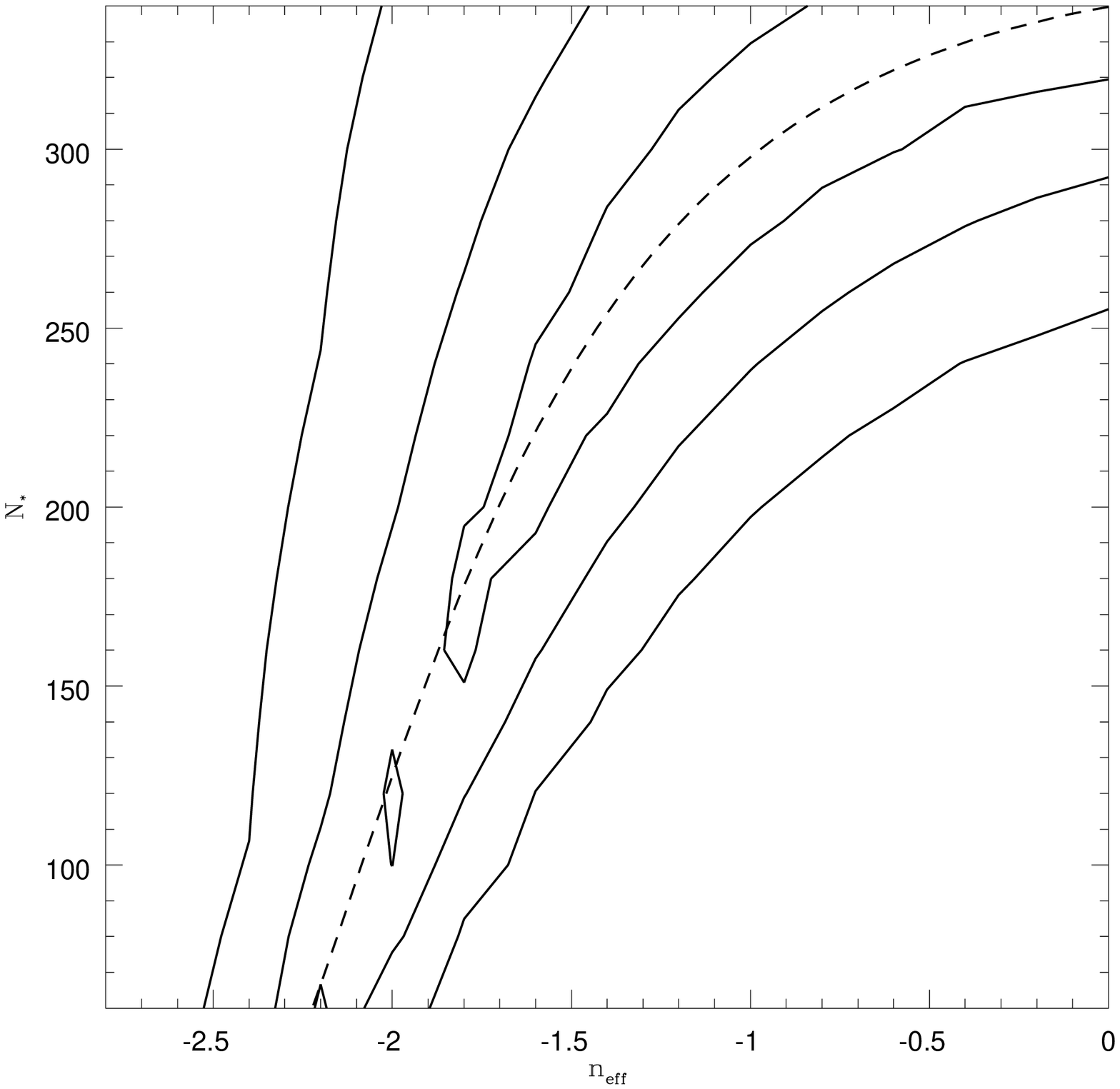}
\caption{The PS model: the contours show
the KS test significance
level $S_{KS}(N_{mem})$ for the comparison between the
observed distribution of the groups and
the PS model described by  $N_{\star}$ and the effective spectral index 
$n_{eff}$.
The outer contours indicate the $99\%$ confidence level, the middle
contours indicate the $50\%$ confidence level and the inner contours
show the $10\%$ confidence level.
The dot-dashed line shows the relation for
data distributed according to a power law (eq.~\ref{nstar})
and with $<N> =300$, $\gamma=2.9$ and $k=1.43$.
The power-law line lies well within the
$50\%$ confidence level. The relation between the group richness $N$
and
mass is $M \propto N^{1.43}$
(see Table 1).}
\end{figure}

\begin{figure}\figurenum{5}
\plotone{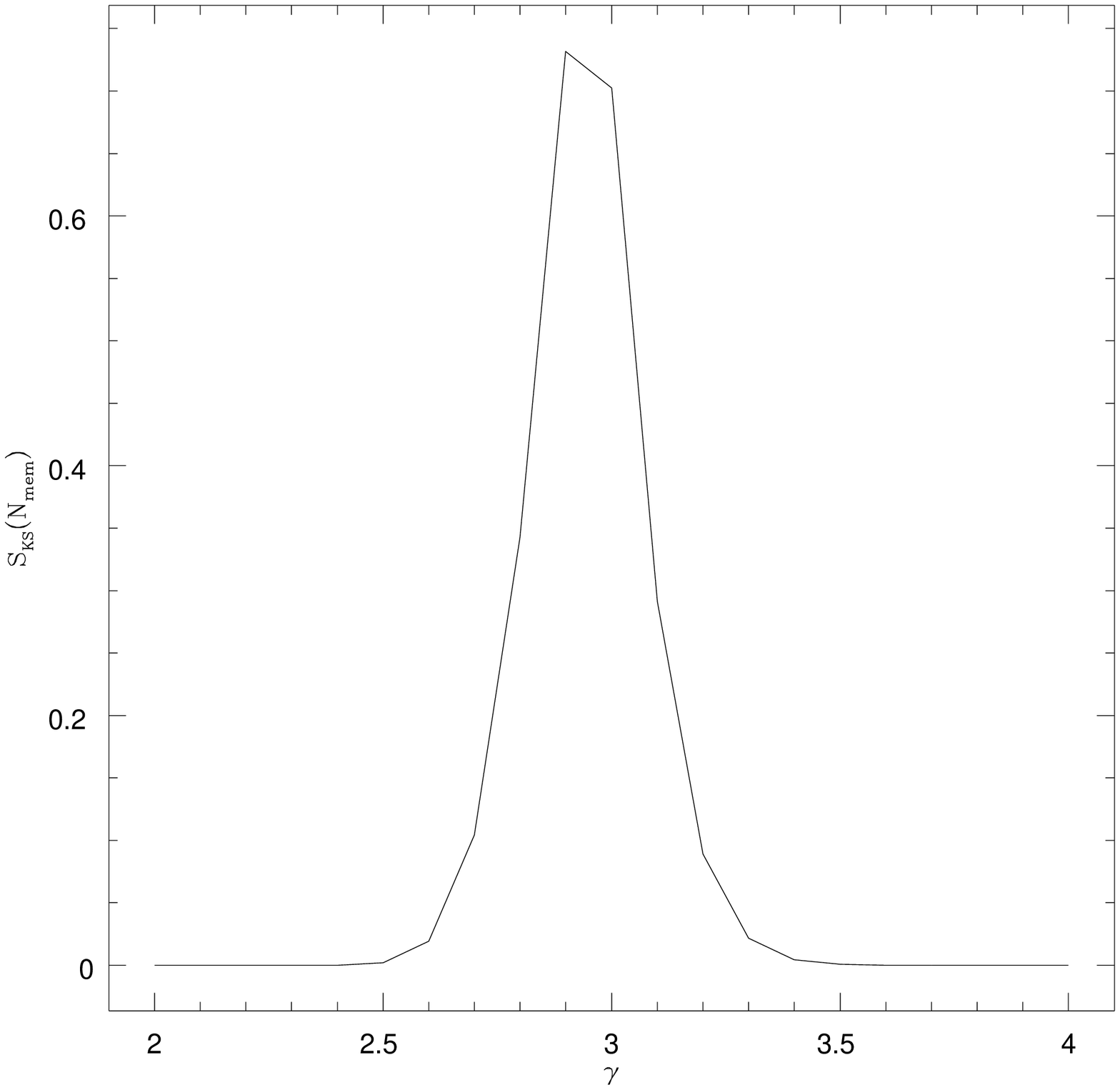}
\caption{The power law model: the KS test significance
level $S_{KS}(N_{mem})$ for the comparison between the model and
observed distribution of the group members.}
\end{figure}

\begin{figure}\figurenum{6}
\plotone{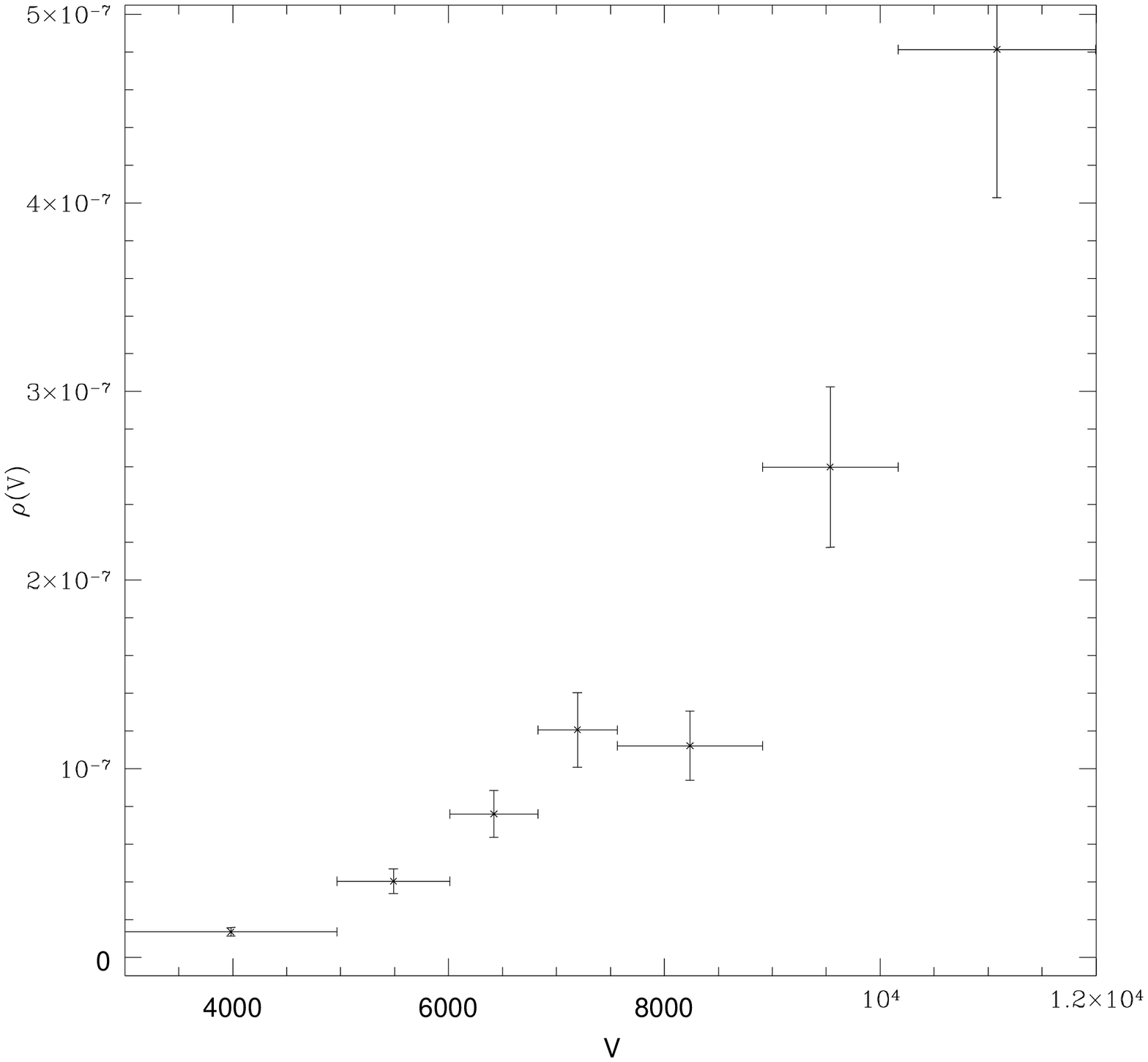}
\caption{[Fig. 6] The density of groups as a function of the radial
distance $\rho(V)$. The function is  the best fit power law
model
for $p(N)$ with $N_{shell}=7$ intervals. The error bars in
$V$
are the width of the bins; for $\rho$ they the
Poisson uncertainty $\delta N_i$ in the number of groups per bin.}
\end{figure}

\begin{figure}\figurenum{7}
\plotone{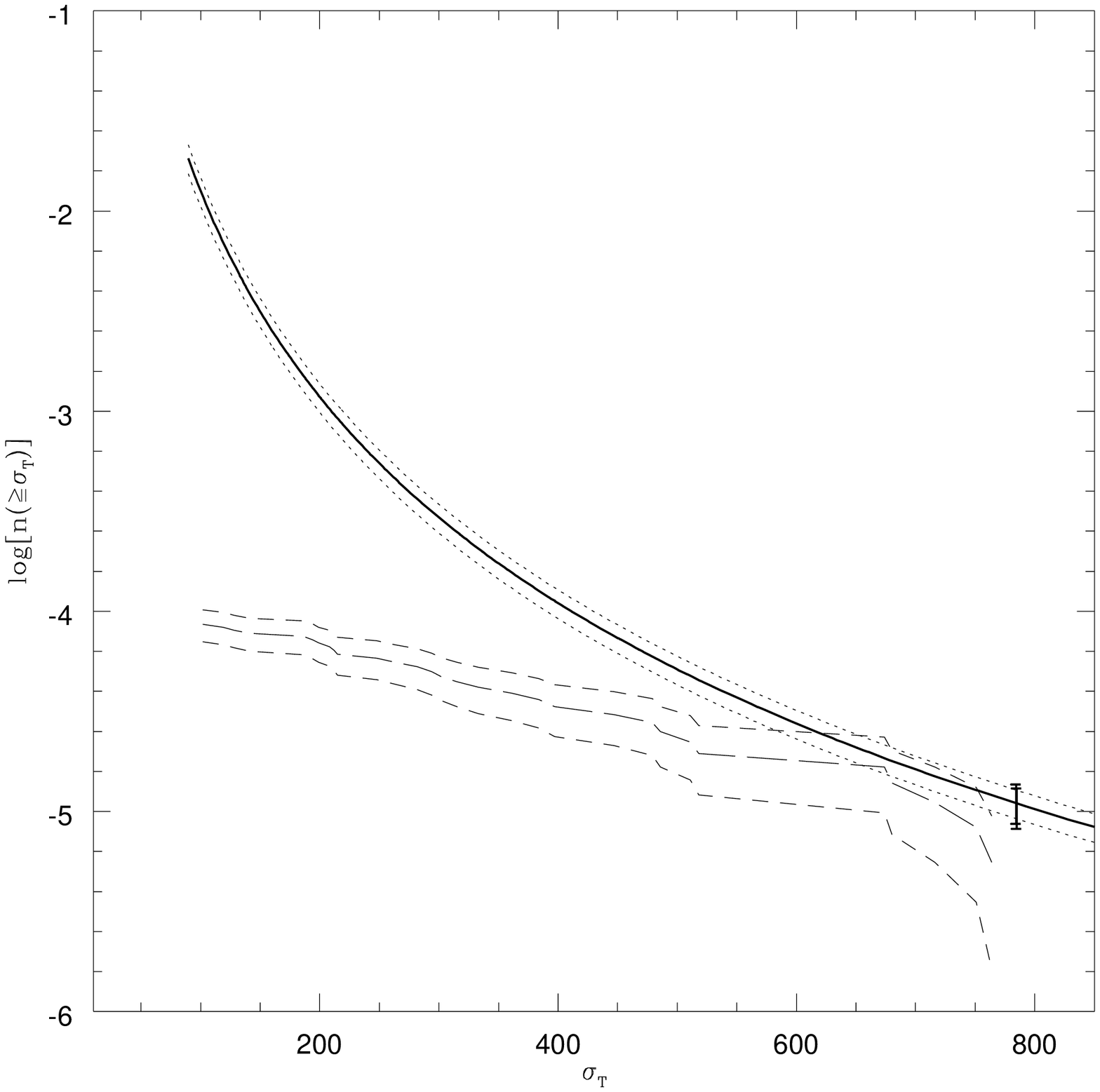}
\caption{The power law model for the
space density of groups with true velocity dispersion $\geq \sigma_{T}$.  The
thin curve shows the power law model (Eq.~\ref{n_st_result}), the dotted
lines show the $95\%$ confidence band for the entire range of 
$\sigma_T$.  The right side vertical bar shows the width of the $95\%$
confidence level for the observed density of robust groups.  The
dashed curves show the data of \citet{zab93} ($\pm$ $1-\sigma$
poissonian error band).}
\end{figure}

\begin{figure}\figurenum{8}
\plotone{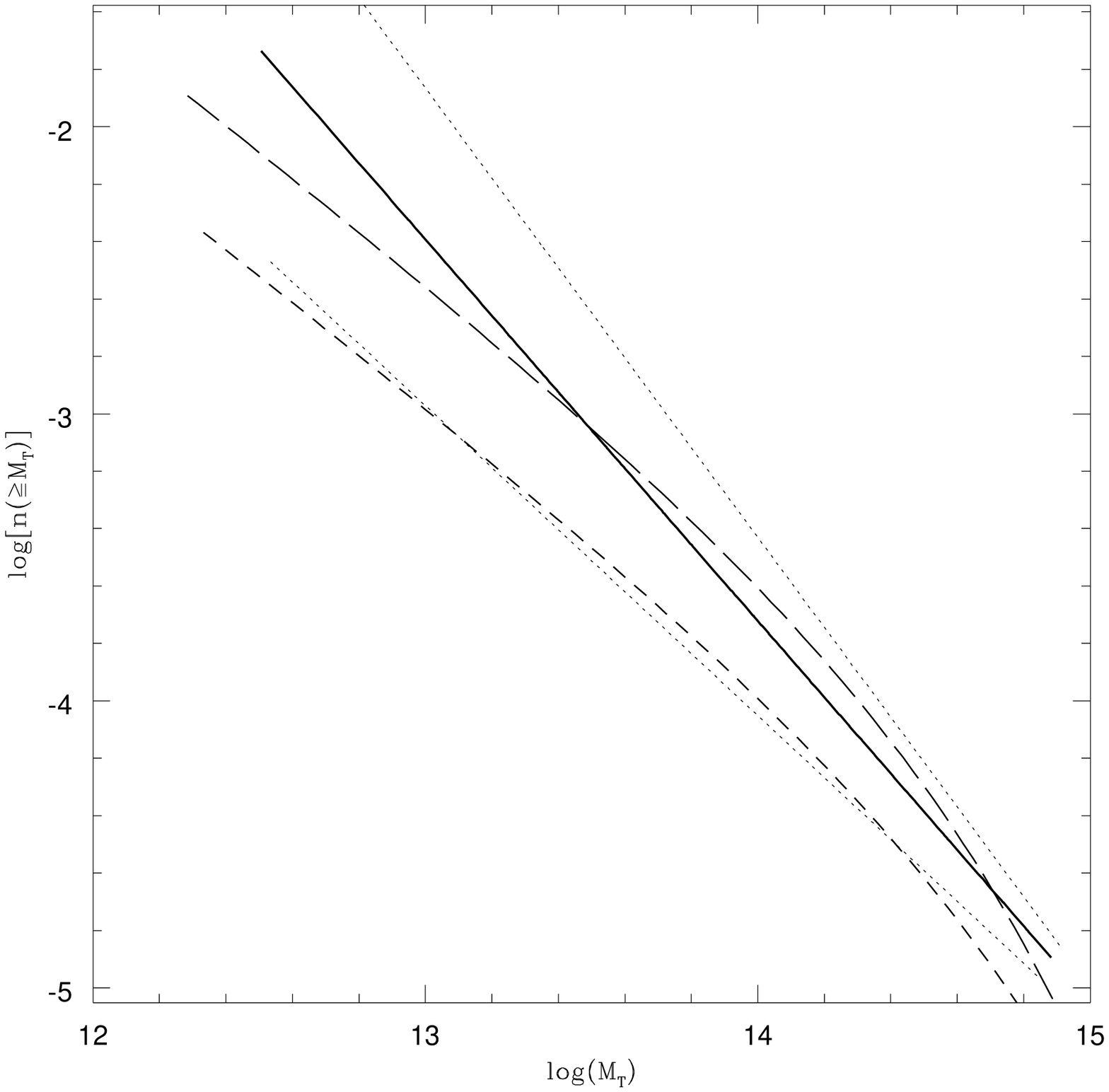}
\caption{The cumulative mass function of groups (thick solid line).
Dotted lines are 95\% confidence levels. The thick short- and long-dashed 
lines
are theoretical predictions \citep{jen01} for $\Lambda$CDM ($\Omega_m =
0.3, \Omega_\Lambda = 0.7, \sigma_8 = 0.9$) and $\tau$CDM with  ($\Omega_m = 1,
\Omega_\Lambda = 0, \sigma_8 = 0.6$) models respectively.}
\end{figure}

\begin{figure}\figurenum{9}
\plotone{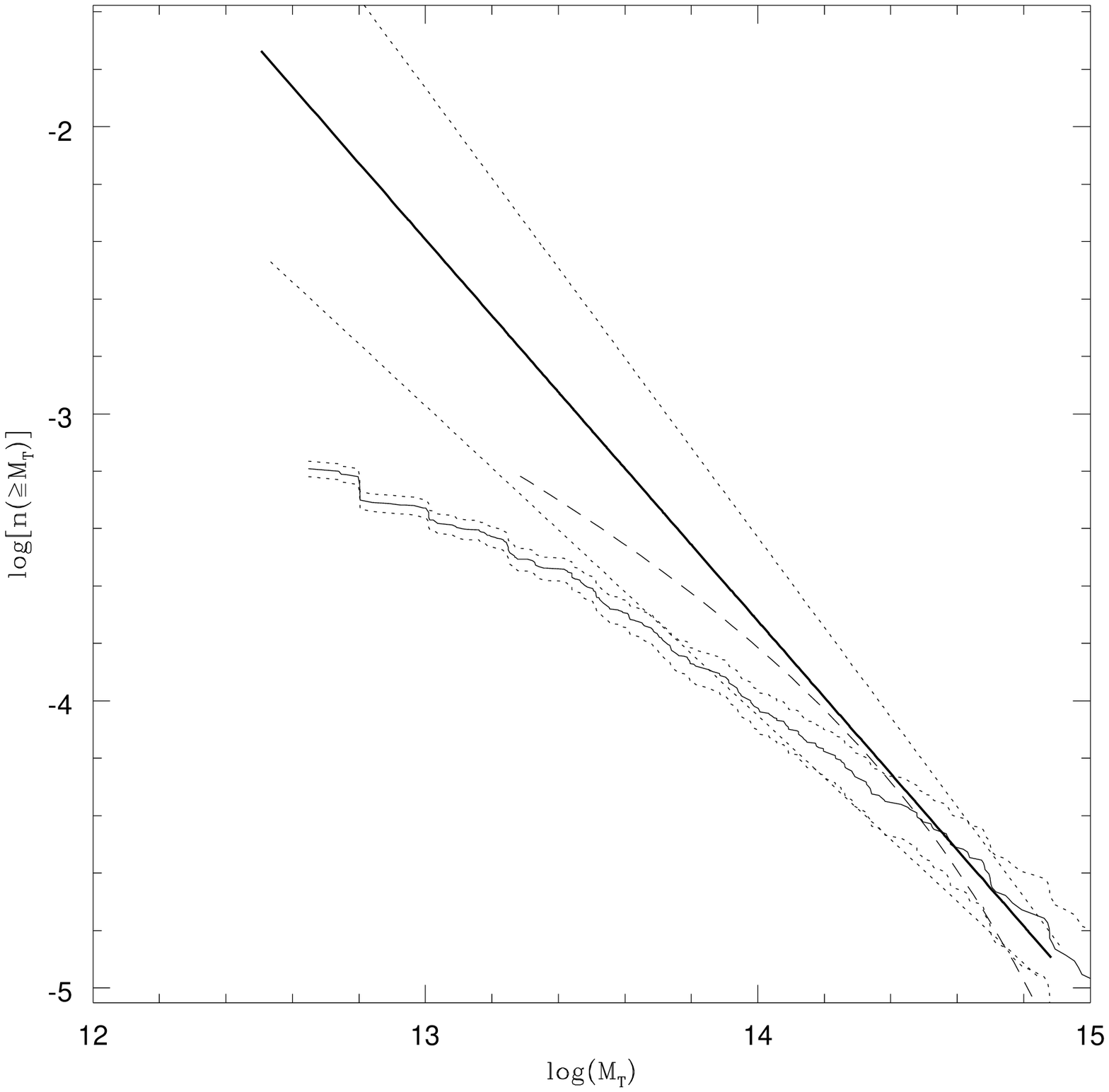}
\caption{The cumulative mass function of groups (thick solid line).
Dotted lines are 95\% confidence levels. The dashed line shows the results
obtained by \citet{gg00}. The thin solid curve is the
observed mass function we obtain assuming constant large scale
density of groups.}
\end{figure}

\begin{figure}\figurenum{10a}
\plotone{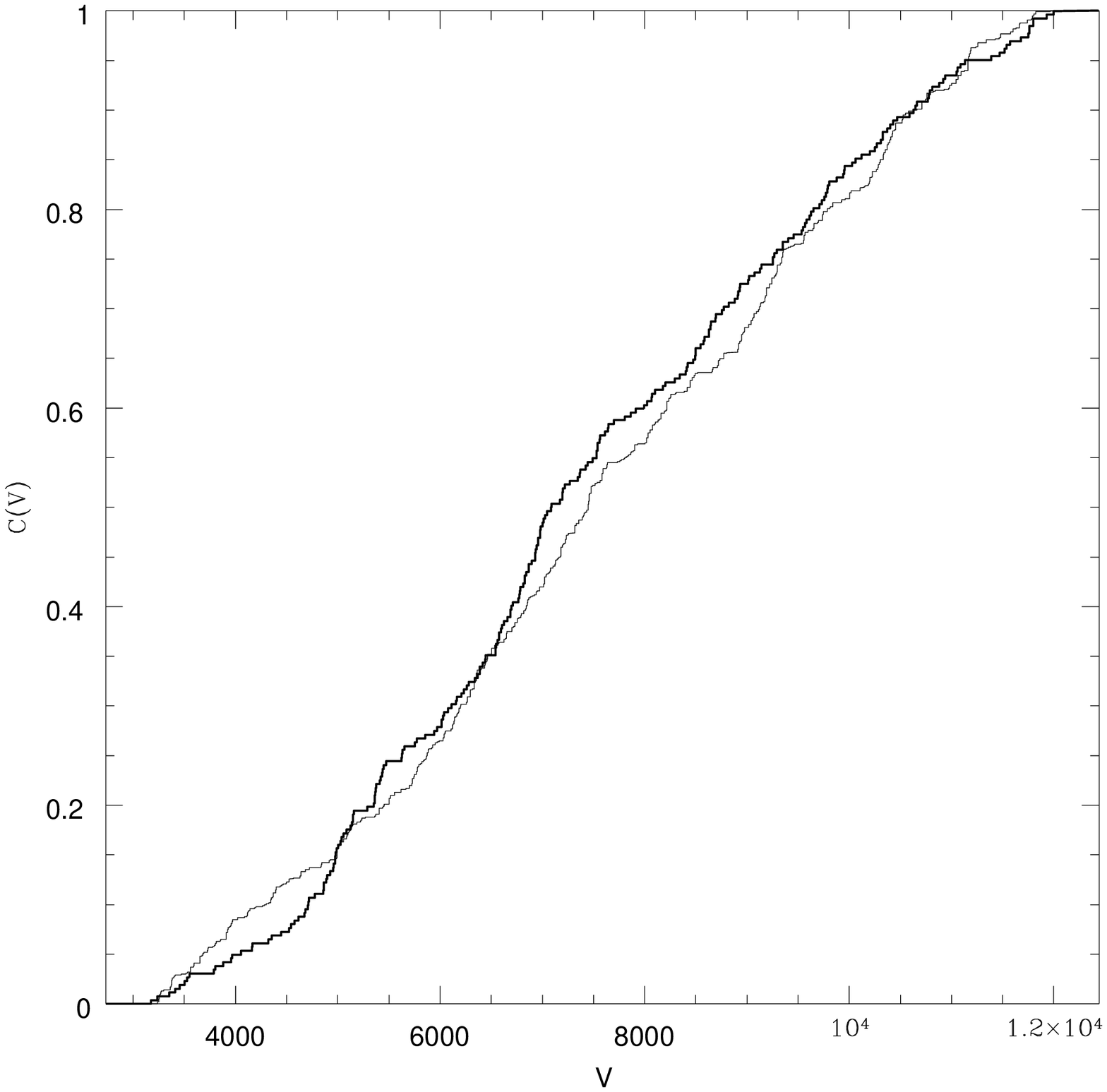}
\caption{[Figure 10a] The Power Law model for the group pdf: the observed
and model cumulative distribution of  radial velocity $V$
of UZCGG groups (thick curve) and a 1000 group Monte Carlo simulation
of our best fit power law model (thin curve).}
\end{figure}

\begin{figure}\figurenum{10b}
\plotone{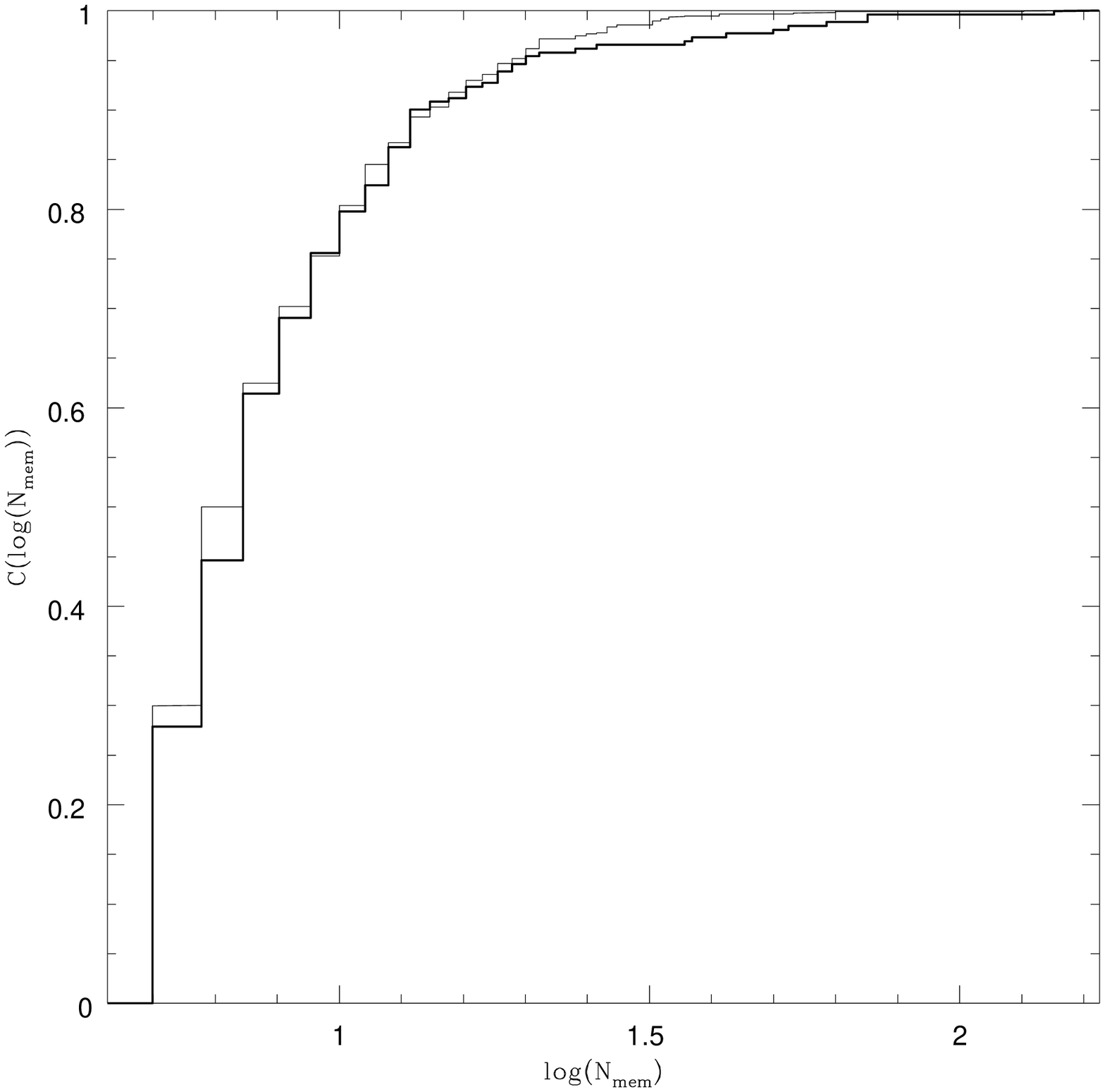}
\caption{[Figure 10b] The power law model for the group pdf: observed
and model cumulative distribution of  $N_{mem}$
 of UZCGG
groups (thick curve) and a 1000 group Monte Carlo simulation
of our best fit power law model (thin curve).]}
\end{figure}

\begin{figure}\figurenum{10c}
\plotone{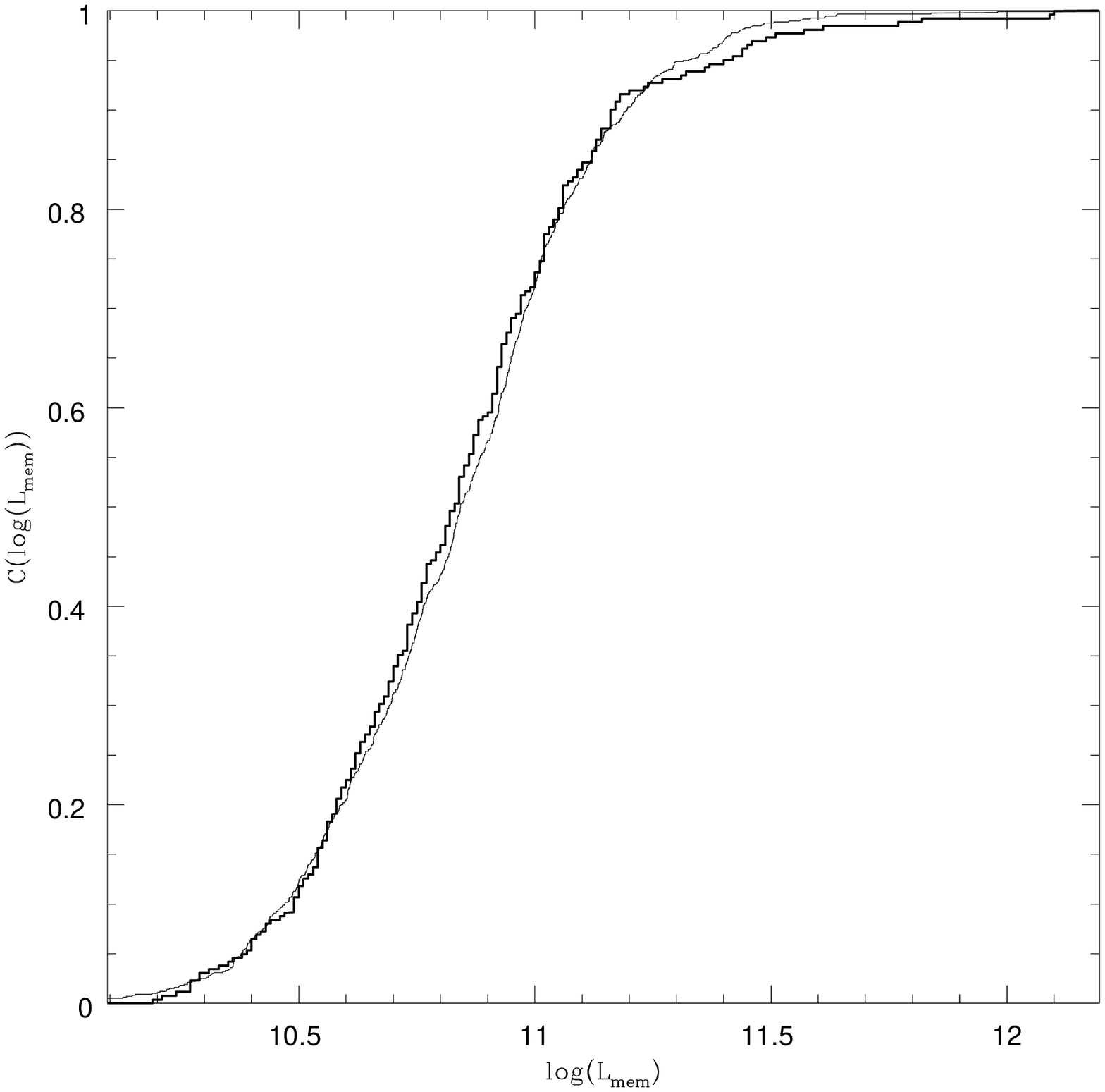}
\caption{[Figure 10c] The power law model for the group pdf: observed
and model cumulative distribution of  $\log(L_{mem})$
 of UZCGG
groups (thick curve) and a 1000 group Monte Carlo simulation
of our best fit power law model (thin curve).]}
\end{figure}

\begin{figure}\figurenum{10d}
\plotone{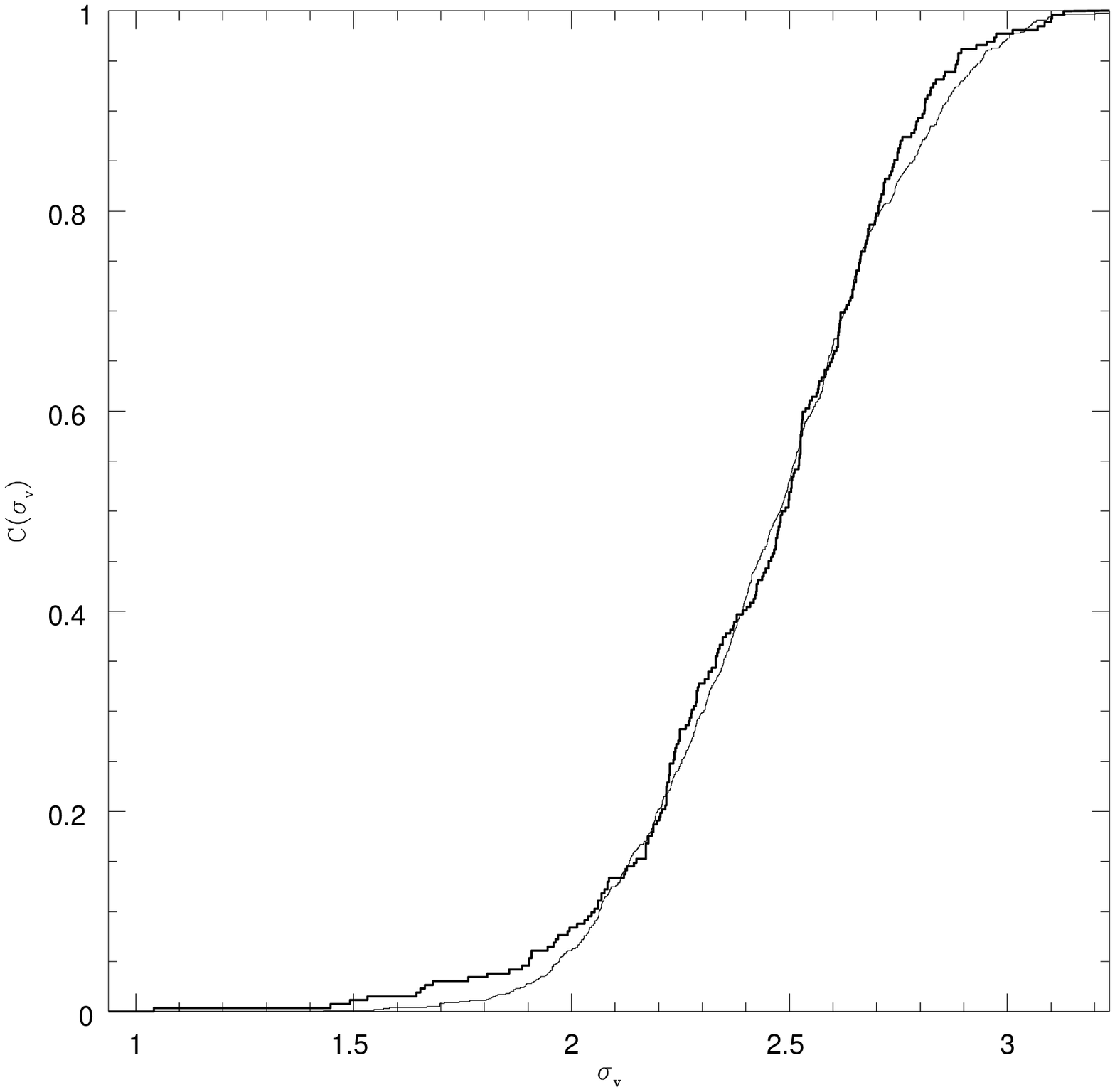}
\caption{[Figure 10d] The power law model for the group pdf: observed
and model cumulative distribution of  $\sigma_v$
 of UZCGG
groups (thick curve) and a 1000 group Monte Carlo simulation
of our best fit power law model (thin curve).]}
\end{figure}

\begin{figure}\figurenum{10e}
\plotone{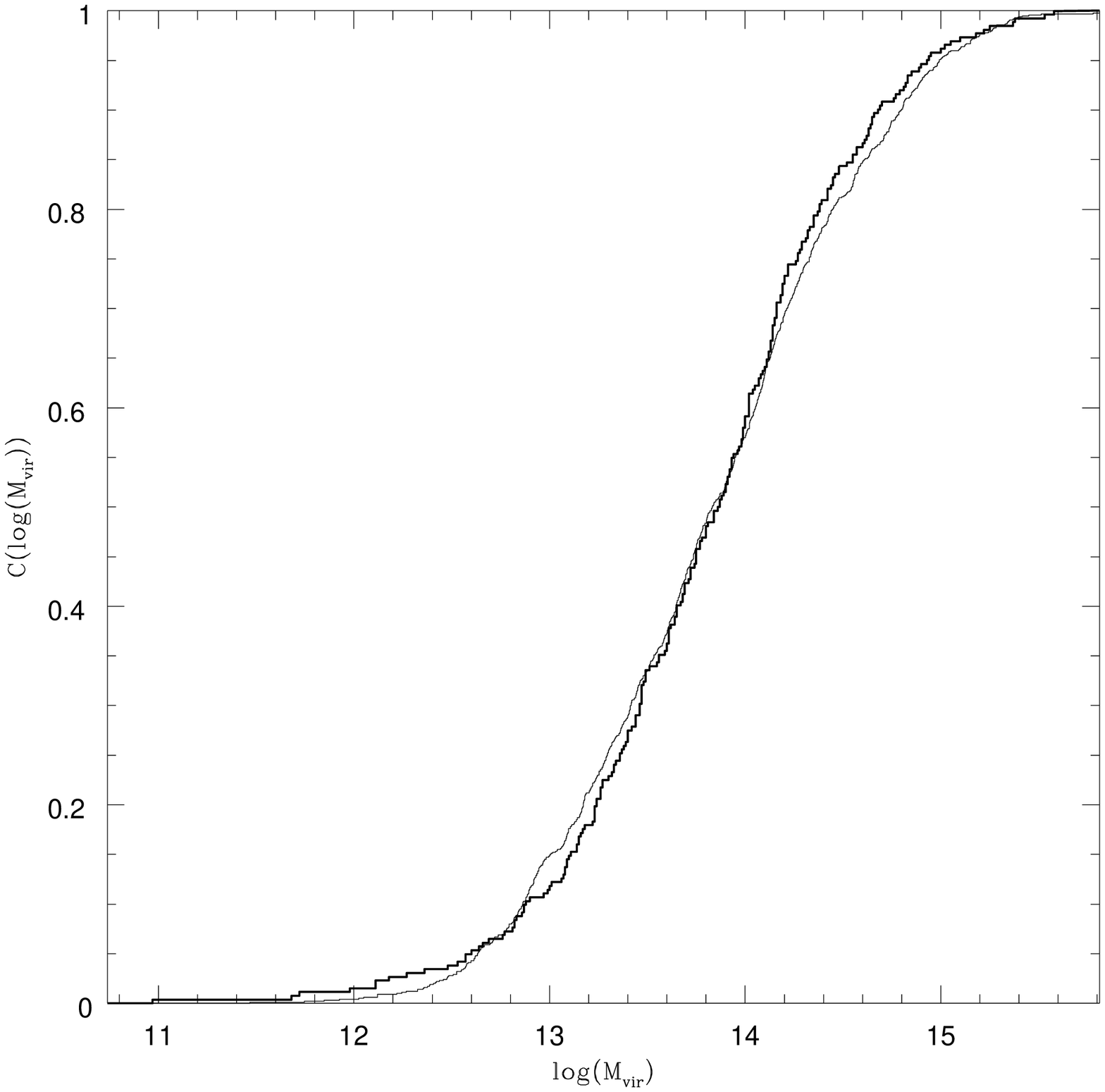}
\caption{[Figure 10e] The power law model for the group pdf: observed
and model cumulative distribution of  virial mass $M_{vir}$
 of UZCGG
groups (thick curve) and a 1000 group Monte Carlo simulation
of our best fit power law model (thin curve).]}
\end{figure}

\clearpage

\begin{deluxetable}{ccc}
\tablecaption{The scaling relations among the main group parameters. \label{tbl-1}}
\tablewidth{0pt}
\tablehead{\colhead{Relation} & \colhead{Slope}   & \colhead{Intercept}  }
\startdata
$\log(M_{T})$ vs. $\log(N)$ & $1.43^{+0.04}_{-0.04}$ &
$10.70^{+0.08}_{-0.09}$
\\
$\log(M_{T})$ vs. $\log(L_{T})$ & $1.44^{+0.06}_{-0.06}$ &
$-2.21^{+0.66}_{-0.71}$ \\
$\log(M_{T})$ vs. $\log(\sigma_{T})$ & $2.58^{+0.07}_{-0.07}$ &
$7.48^{+0.16}_{-0.18} $\\
$\log(L_{T})$ vs. $\log(\sigma_{T})$ & $1.80^{+0.05}_{-0.04}$ & $6.71^
{+0.11}_{-0.11}$\\
$\log(N)$ vs. $\log(R_{vir})$ & $2.00^{+1.54}_{-0.88}$ & $2.11^
{+0.06}_{-0.02}$ \\
$\log(\sigma_{T})$ vs. $\log(R_{vir})$ & $1.18^{+0.88}_{-0.45}$ & $2.42^
{+0.03}_{-0.08}$

 \enddata

\end{deluxetable}

\end{document}